\documentclass[twocolumn,superscriptaddress,amsmath,amssymb,aps,prb]{revtex4-2}

\pdfoutput=1

\usepackage[utf8]{inputenc}

\usepackage{amsmath}
\usepackage{amsfonts}
\usepackage{amsthm}
\usepackage{mathtools} 

\usepackage[hyperindex,breaklinks]{hyperref}
\usepackage{url}

\usepackage{gensymb}
\usepackage{graphicx}
\usepackage{dcolumn}
\usepackage{array}
\usepackage{bm}
\usepackage{upgreek}

\newcommand{\vcr}[1]{\boldsymbol{\mathrm{#1}}}
\newcommand{\vcrg}[1]{\boldsymbol{#1}}

\newcommand{\Abs}[1]{\left\vert #1 \right\vert}

\newcommand{\SqB}[1]{ \left[ #1 \right]}
\newcommand{\RnB}[1]{ \left( #1 \right)}
\newcommand{\CrB}[1]{ \left\{ #1 \right\}}

\newcommand{\myFig}[6]{ %
\begin{figure}[htb]
\begin{center}
\includegraphics[width=#1\columnwidth,height=#2\columnwidth,clip=true,keepaspectratio=#3]{#4}
\caption{#5} \vspace{-0.5cm} \label{#6} 
\end{center}
\end{figure}} 

\begin{document}
\title{Spin transfer torque in Mn$_3$Ga-based ferrimagnetic tunnel junctions from first principles}
\author{Maria Stamenova}\email[Contact email address: ]{stamenom@tcd.ie}
\affiliation{School of Physics and CRANN, Trinity College, Dublin 2, Ireland} 
\author{Plamen Stamenov}
\affiliation{School of Physics and CRANN, Trinity College, Dublin 2, Ireland}
\author{Farzad Mahfouzi}
\affiliation{Department of Physics and Astronomy, California State University, Northridge, California 91330, USA}
\author{Quilong Sun}
\affiliation{Department of Physics and Astronomy, California State University, Northridge, California 91330, USA}
\author{Nicholas Kioussis}
\affiliation{Department of Physics and Astronomy, California State University, Northridge, California 91330, USA}
\author{Stefano Sanvito} 
\affiliation{School of Physics and CRANN, Trinity College, Dublin 2, Ireland} 

\begin{abstract}
We report on first-principles calculations of spin-transfer torque (STT) in epitaxial magnetic tunnel junctions (MTJs) based on ferrimagnetic tetragonal Mn$_3$Ga electrodes, both as analyzer in an Fe/MgO stack, and also in an analogous stack with a second Mn$_3$Ga electrode (instead of Fe) as polarizer. Solving the ballistic transport problem (NEGF + DFT) for the nonequilibrium spin density in a scattering region extended to over 7.6 nm into the Mn$_3$Ga electrode, we find long-range spatial oscillations of the STT decaying on a length scale of a few tens of angstroms, both in the linear response regime and for finite bias. The oscillatory behavior of the STT in Mn$_3$Ga is robust against variations in the stack geometry (e.g., the barrier thickness and the interface spacing) and the applied bias voltage, which may affect the phase and the amplitude of the spatial oscillation, but the wave number is only responsive to variations in the longitudinal lattice constant of Mn$_3$Ga (for fixed in-plane geometry) without being commensurate with the lattice. Our interpretation of the long-range STT oscillations is based on the bulk electronic structure of Mn$_3$Ga, taking also into account the spin-filtering properties of the MgO barrier. Comparison to a fully Mn$_3$Ga-based stack shows similar STT oscillations, but a significant enhancement of both the TMR effect at the Fermi level and the STT at the interface, due to resonant tunneling for the mirror-symmetric junction with thinner barrier (three monoatomic layers). From the calculated energy dependence of the spin-polarized transmissions at 0\,V, we anticipate asymmetric or symmetric TMR as a function of the applied bias voltage for the Fe-based and the all-Mn$_3$Ga stacks, respectively, which also both exhibit a sign change below 1\,V. In the latter (symmetric) case we expect a TMR peak at zero, which is larger for the thinner barriers because of a spin-polarized resonant tunneling contribution.   
\end{abstract}


\maketitle
\section {Introduction} \label{sect:intro}

The Fe/MgO-based magnetic tunnel junctions (MTJs) are the backbone of modern spintronics and the idealised crystalline Fe(100)/MgO/Fe MTJ is the theoretical proxy system for the locally structurally-coherent CoFeB/MgO/CoFeB MTJs. It is the former structure, where the spin-filtering tunneling-magnetoresistance (TMR) effect was predicted theoretically\cite{Butler2001} some 20 years ago and soon after demonstrated experimentally\cite{Parkin2004,Yuasa2004}. In essence, the TMR effect, which exploits the difference in resistivity between parallelly and anti-parallelly aligned magnetic layers sandwiching an insulator, in these MTJs, is due the special symmetry-driven spin-filtering of the Fe(100)/MgO composite. In a few atomic mono-layers (MLs) of MgO, the transmission of the minority spin carriers emanating from Fe(100) is almost completely eliminated and theoretically the TMR effect, in an ideal Fe/MgO/Fe MTJ, can reach several thousands of percent\cite{Butler2008,Rungger2009,Rungger2018}. The TMR effect combined with the possibility of switching or exciting precession in the free magnetic layer by current, due to the spin-transfer torque (STT), makes the Fe/MgO MTJs suitable functional components in magnetic memory elements or high frequency generators \cite{Houssameddine2007,Heindl2019,Hirohata2020}. Those applications for MTJs require the optimisation of certain magnetic properties.              

The combination of high spin-polarisation, low Gilbert damping and large anisotropy is highly desirable for the scalability of spintronic applications like high-density spin-transfer torque memory (STT-MRAM) or spintronic oscillators and detectors in the THz range. Mn-Ga alloys have been studied for magnetisation dynamics applications because of their relatively high anisotropy, for a low-$Z$ material, and indeed found to exhibit low Gilbert damping coefficients \cite{Mizukami2011} as well. In addition, the tetragonal Heusler DO$_{22}$ form of Mn$_3$Ga exhibits a low-moment, ferrimagnetic order and a high spin polarization\cite{Rode2013}. A further reason for studying this system, in particular, is its similarity with the prototype fully-compensated half-metallic Mn$_x$Ru$_{1-x}$Ga compound, a very topical material exhibiting high spin-polarization, low damping and strong perpendicular anisotropy but as of site-disorder -- rather difficult to simulate. For instance, recently current-induced switching with interfacial spin–orbit torque has been demonstrated for ultrathin films of the latter Heusler compound, interfaced with Pt\cite{Finley2018}. Similarly, the DO$_{22}$ structure of Mn$_3$Ga is ferrimagnetic,  also featuring two antiferromagnetically-coupled Mn sublattices -- one formed of Mn atoms, labeled as Mn$_\mathrm{I}$ in the $2b$ Wyckoff positions, e.g. $(0,0,1/2)$, forming Mn$_\mathrm{I}$-Ga planes; and the other sublattice of Mn$_\mathrm{II}$ atoms in the $4d$ positions, like $(0,1/2,1/4)$, forming Mn$_\mathrm{II}$-Mn$_\mathrm{II}$ planes [see Fig. \ref{fig01}(a)].    

STT-driven thin Mn$_3$Ga free-layers, as parts of MTJ stacks, are interesting on their own for the construction of STT-driven oscillators, because of their comparatively large effective anisotropy and corresponding ferromagnetic resonance frequencies of even the in-phase modes. The observed resonance frequencies of stand-alone films vary from about 0.17 THz to above 0.35 THz, for thicknesses in the range 4 - 15\,nm, respectively, with the emission bandwidth decreasing monotonically as a function of increasing thickness from above 40 GHz to below 25 GHz.\cite{Ilyakov2019,Awari2016} While coherent low-THz range emission is still to be demonstrated from this type of moderate spin polarisation ($P\sim 45 \%$) electrode under current excitation, within nano-pillar structures, the nature and magnitude of the STT and theoretical maximal efficiencies, with which the in-phase and out-of-phase resonance modes can be excited, remain open questions.

We consider mesoscopic junctions in which the Mn$_3$Ga film is grown on top of the Fe(100)/MgO stack in the longitudinal $z$ direction [see Fig. \ref{fig01}(a)], while there are periodic boundary conditions in the $x$-$y$ plane. The open-boundary conditions are applied at the two ends of the scattering region (SR) of the stack, depicted in Fig. \ref{fig01}(a), via the non-equilibrium Green's function (NEGF) method, as implemented in the Smeagol code.\cite{Alex04} In practice, there are two semi-infinite crystalline leads of bcc Fe and DO$_{22}$ tetragonal Mn$_3$Ga attached to the left and to the right end of the SR, respectively. Thus constructed, the stack is laterally commensurate to the lattice of bcc Fe with its lattice constant $a_\mathrm{Fe}=2.866$\,\AA, which is, rotated by 45$\degree$ with respect to the cubic MgO lattice. Hence, we consider the tetragonal Mn$_3$Ga cast into the Fe-dictated in-plane lattice constants $a=b=\sqrt{2}a_\mathrm{Fe}=4.053$\,\AA\, in the lateral directions. This leads to a $+3.7$\,\% tensile bi-axial strain in Mn$_3$Ga (from the experimental lateral lattice constant of 3.91\,\AA\cite{Rode2013}) and a $-3.8$\,\% compressive bi-axial strain in the MgO. Geometry relaxations, at the level of the local spin-density approximation (LSDA) to the exchange-correlation functional \cite{Perdew1981} and constrained to the longitudinal direction only, have resulted in a significant ($\sim$15\,\%) compression of the Mn$_3$Ga slab with respect to the experimental value $c_\mathrm{exp}=7.1$\,\AA. \cite{Rode2013} This, in turn, leads to unrealistically small values of the local magnetic moments. These shortcomings of the LSDA geometry for Heusler alloys are known and typically the GGA (PBE) are used.\cite{Rode2013,Zic2016} Here we are, however, limited to the LSDA for our transport calculations of multi-layered junctions with non-collinear spin alignments. As a compromise, the value of $c$ has been chosen such that, within the LSDA, the calculated atomically-projected local spins (as per Mulliken population analysis) are within the experimental ranges for the spins of the Mn atoms, obtained by different measuring techniques (XMCD or neutron diffraction),\cite{Rode2013} namely $s_\mathrm{Mn_I} \in \RnB{3.2, 3.7}\mu_\mathrm{B}$ and $s_\mathrm{Mn_{II}} \in \RnB{-2.1, -2.7 }\mu_\mathrm{B}$. Note that Mn$_\mathrm{I}$ is in the planes with Ga, while Mn$_\mathrm{II}$ forms Mn$_\mathrm{II}$-Mn$_\mathrm{II}$ planes perpendicular to the direction of the transport [see Fig. \ref{fig01}(b)]. Furthermore, our strategy has been, instead of limiting ourselves to a single albeit accurate geometry optimisation beyond LSDA, to explore a range of structural parameters in the $z$-direction, i.e. values of $c$ and $d$ -- the distance at the Mn$_3$Ga-MgO interface (the other side of the junction, the Fe/MgO, is as in Ref. \onlinecite{Rungger2009}). For our main representative structure we have chosen $c=6.6$\,\AA, which results in magnetic moments for the two types of Mn atoms in the experimental value ranges [see Fig. \ref{fig01}(c) for the atomically-resolved (Mulliken population analysis) spin values]. We also consider the two different possible Mn$_3$Ga terminations on the $(001)$ interface with MgO, but the representative case (used as reference throughout the paper, unless stated otherwise), depicted in Fig. \ref{fig01}(a), features a Mn$_\mathrm{I}$-Ga-plane termination (shown in the zoomed-in inset). In this termination Mn$_\mathrm{I}$ atoms are placed on top of the oxygen atoms. In our representative case the interfacial Mn-O spacing $d=2.215$\,\AA\, corresponds approximately to half lattice constant of bulk $\alpha$-MnO.\footnote{From the International Centre for Diffraction Data, \url{https://www.icdd.com/}.} Note, that this value of interlayer distance at the interface agrees well with the value of 2.265\,\AA\, that we found using VASP calculations with the PBE exchange correlation functional for the Mn$_3$Ga/MgO slab with Mn$_\mathrm{I}$-Ga terminated interface.~\cite{Sun2020} 

\myFig{1}{1}{true}{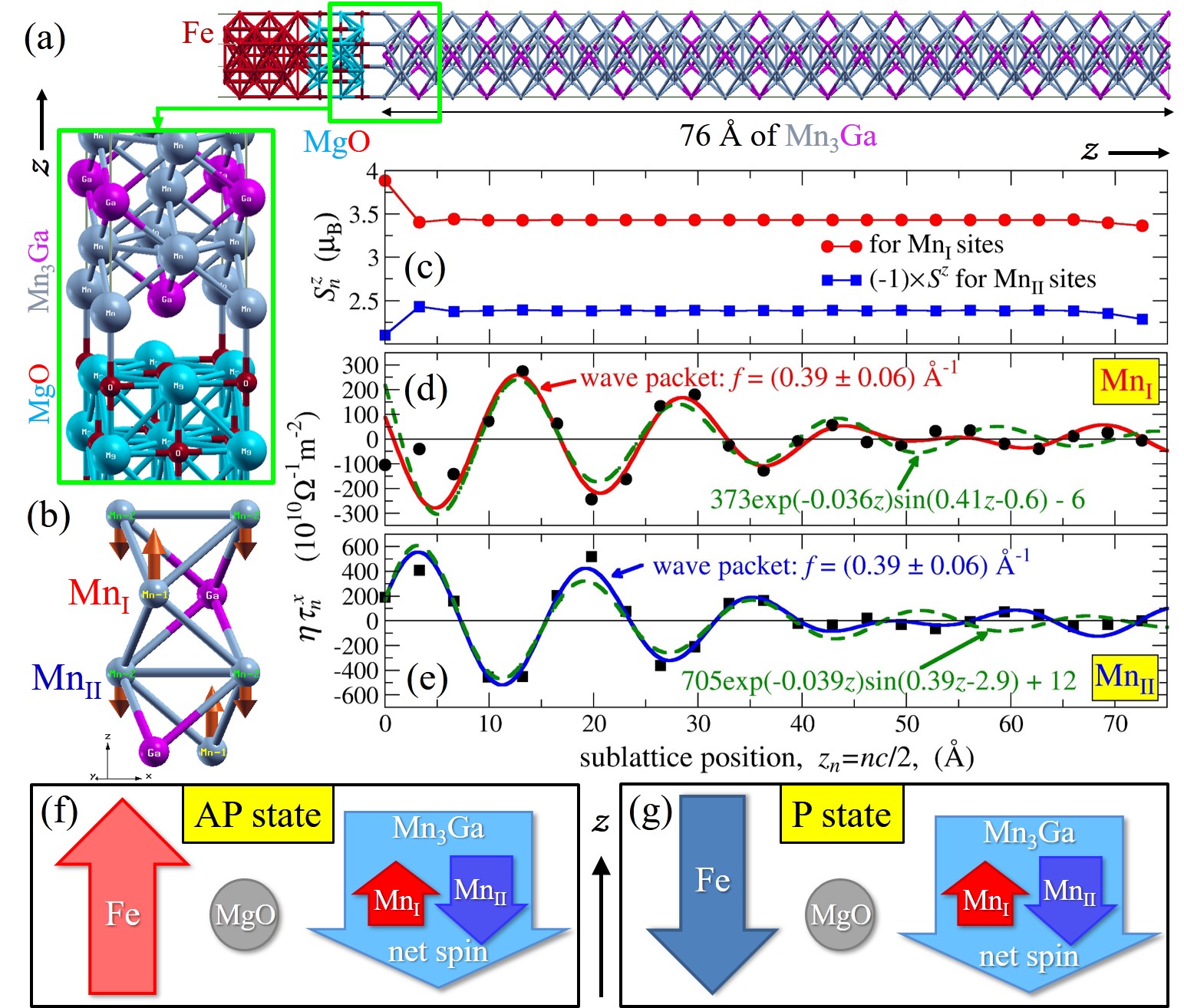}{(Color online) (a) Schematic of the Fe/MgO/Mn$_3$Ga junction with a close-up of the Mn$_3$Ga/MgO interface (in the green rectangle), depicting the main geometry investigated with Mn$_\mathrm{I}$-Ga termination. (b) The tetragonal unit cell of Mn$_3$Ga showing the directions of the spins of the anti-ferromagnetically (AFM) coupled Mn$_\mathrm{I}$ and Mn$_\mathrm{II}$ sublattices. (c) The magnitudes of the local spins at Mn$_\mathrm{I}$ and Mn$_\mathrm{II}$ sites in the junction starting from the MgO interface, as depicted in the schematic above. (d,e) Corresponding calculated atomically-resolved in-plane STTk, as described by Eq. (\ref{eq:STTk}) for the Mn$_\mathrm{I}$ and Mn$_\mathrm{II}$ sublattices, respectively, for the case in which the moments on the Mn are along $z$, while the moment of the Fe lead is along $x$. The coefficient $\eta\equiv e/(\mu_\mathrm{B}A)$, where $A$ is the transverse area of the junction, is used throughout the paper to convert our computed STT (or STTk) to units $\Omega^{-1}\mathrm{m}^{-2}$. (f,g) define what we refer as the anti-parallel (AP) and parallel (P) spin state of the junction, respectively.}{fig01}

As far as the magnetic state is concerned, there are two possible collinear-spin configurations of the junction. We disregard the spin-orbit interaction, hence there is no coupling between the spatial orientations of the spins and the geometry of the junction. For definiteness, and in view of the expected perpendicular anisotropy in these junctions \footnote{The junctions targeted for experimental comparisons are primarily based on ultra-thin CoFeB polarizers, which indeed have a well-developed perpendicular anisotropy, provided by the spin-orbit coupling at the interface of Fe and MgO. Here we neglect the small differences in polarisation and anisotropy between Fe and CoFe.}, our quantization axis is oriented along the direction of the transport $z$, see Fig. \ref{fig01}(a,b). We define two possible collinear states of the junction: a parallel (P) state in which the net moment  of Mn$_3$Ga is parallel to Fe, and AP when it is anti-parallel to that of Fe. Note that, as the net spin is parallel to Mn$_\mathrm{II}$ and anti-parallel to Mn$_\mathrm{I}$ this means, for our preferred Mn$_\mathrm{I}$-Ga termination, that the Mn$_\mathrm{I}$ spin at the interface opposes the Fe spin in the P state and is parallel to it in the AP state [Fig. \ref{fig01}(f,g)]. For our steady-state transport calculations of STT, the typical non-collinear state we consider is with the Fe moment rotated along the $x$-axis so that there is a 90$\degree$ misalignment between spins in the two electrodes.                     

The paper is then organised as follows. In the next Section \ref{sect:theory} we outline the formalism used for the calculation of the linear response ST-torkance (STTk) and then we present some technical insights of the calculated atomically-resolved STTk for the Mn$_3$Ga layer in the self-consistently described scattering region. In Section \ref{sect:barrier}, we focus on the effects of the interface and the barrier geometries on the in-plane STTk. In Section \ref{sect:lattice} we examine the effect of the bulk lattice parameters (in particular, the long axis lattice constant $c$) of Mn$_3$Ga and elucidate the origin of the observed long-range spatial oscillation of the STTk. In Section \ref{sect:finbias} we discuss the spin-polarised transmission at equilibrium, decomposed over the transverse two-dimensional Brillouin zone (2D BZ) or as a function of energy.  Based on that, we then evaluate the TMR near equilibrium and draw predictions for its asymmetric bias dependence in the range -1\,V to 1\,V. In Section \ref{sect:last} we look at a modified junction, where we replace the Fe lead with Mn$_3$Ga, hence constructing a mirror-symmetric all-ferrimagnetic MTJ (FiMTJ), for which we compare analogously-calculated STT and TMR properties to the Fe-based junction. Then we conclude with a discussion and comparison to existing experimental data on related MnGa-based tunnel junctions. \cite{Borisov2016}

\section{Linear response STT} \label{sect:theory}

We calculate the linear response STT using the method from Ref. [\onlinecite{Stamenova2016}], which is described in greater detail in Ref. [\onlinecite{Rungger2018}] and implemented in the Smeagol code\cite{Alex04}. In this regime a small bias voltage, $\delta V_\mathrm{b}$, is applied across the junction. Then the transport part of the density matrix induced by $\delta V_\mathrm{b}$ is defined as
\begin{equation}\label{eq:rho}
\rho_\mathrm{tr} (\delta V_\mathrm{b}) \approx \left.\frac{\partial \rho (V_\mathrm{b})}{\partial V_\mathrm{b}}\right\vert_{V_\mathrm{b}=0} \, \delta V_\mathrm{b}.
\end{equation} 
and the so-called spin-transfer torkance (STTk) acting on an atomic site $n$, $\vcrg{\tau}_n$, is defined as
\begin{equation}
\vcrg{\uptau}_n = \frac{\delta\vcr{T}_n}{\delta V_\mathrm{b}}
\approx 2 \sum_{\alpha \in n}\sum_{\beta=1}^{N_\beta}\mathrm{Re}\left[\frac{\partial\vcrg{\uprho}_{\mathrm{tr},\beta\alpha}}{\partial V_\mathrm{b}}\times\vcr{H}[\rho_\mathrm{cond}]_{\alpha\beta}\right],
\label{eq:STTk}
\end{equation}
where $\alpha,\beta$ replace the full set of quantum numbers indexing all the orbitals in the local basis set, as implemented in the Siesta code \cite{Soler2002}, $H_{\alpha\beta}=H_{0,\alpha\beta} \vcr{1} +\vcr{H}_{\alpha\beta}\cdot\vcrg{\upsigma}$ is the LSDA Kohn-Sham Hamiltonian of the SR, with $\CrB{\vcrg{1},\vcrg{\upsigma}}$ -- the full set of Pauli matrices, including identity. The density matrix from Eq. (\ref{eq:rho}) is decomposed similarly: $\rho_{\alpha\beta}=\rho_{0,\alpha\beta} \vcr{1} +\vcrg{\uprho}_{\alpha\beta}\cdot\vcrg{\upsigma}$. Note that the integration over the transverse 2D BZ is implicit in Eq. (\ref{eq:STTk}). The required derivative of the density matrix with respect to the bias voltage, within NEGF, is calculated in the linear response regime for each k-point $\vcr{k}\equiv\vcr{k}_\perp \in \Omega_\mathrm{BZ}$ in the 2D BZ surface:
\begin{equation}\label{eq:kdecomp}
\frac{\partial \rho_{\mathrm{tr},\vcr{k}}}{\partial V_\mathrm{b}}\! =\! \frac{1}{4\pi} G_{\vcr{k}}(E_\mathrm{F})\! \left.\SqB{\Gamma_{\vcr{k},\mathrm{L}}(E_\mathrm{F}) - \Gamma_{\vcr{k},\mathrm{R}}(E_\mathrm{F})} G^\dagger_{\vcr{k}}(E_\mathrm{F})\right\vert_{V_\mathrm{b}=0}\,.
\end{equation}
It is assumed that the Green's function of the scattering region, $G(E)$, and the $\Gamma_\mathrm{L(R)}(E)$ matrices, which couple it to the left(right) lead, are slowly-varying functions around the Fermi level, $E_\mathrm{F}$. It is also assumed that the bias drop in the junction is symmetric, i.e. the chemical potentials in the left (right) lead shift by $\pm V_\mathrm{b}/2$ with respect to the equilibrium.

The calculated atomically-resolved in-plane STTk, $\tau^{x}_n$, in Mn$_3$Ga for our representative Fe(100)/MgO/Mn$_3$Ga junction with 45 MLs of Mn$_3$Ga in the SR and Mn$_\mathrm{I}$-Ga terminated interface, is shown in Fig.~\ref{fig01}d,e for the two Mn sublattices, starting from the MgO interface. The net magnetic moment\footnote{Note that we use 'spins' and 'magnetic moments' interchangeably throughout the paper.} in Mn$_3$Ga is oriented along $-z$ (Mn$_\mathrm{I}$ at the interface points in $z$-direction), while the moment of Fe is along $x$. In contrast to the anticipated exponential decay of the STT from the insulating barrier in conventional MTJs, here $\tau^{x}_n (z)$ is showing a slow oscillatory decay for both magnetic sublattices over many MLs of Mn$_3$Ga. The atomically-resolved in-plane STTk in Fig. \ref{fig01} (d,e) is fitted to a beating sine wave, which is one of many possible fitting functions. A decaying sine wave is also a possibility and the size of the data set does not allow to discriminate between those fitting functions. However, we observe that an exponential decay is apparently not as good a fit to the data (the green dashed line), in line with the, so called, {\it spatial precession} behaviour of STT, identified from basic scattering theory principles in a free electron model in Ref. [\onlinecite{Stiles2002}]. We are restricted by the feasibility of the calculation to further extend the self-consistent SR for a more meaningful (quantitatively) non-linear fit. The fitted carrier wavenumber ($2\pi/\lambda$) value is about 0.38\,\AA$^{-1}$, which corresponds to a period of oscillation of some 10 MLs of Mn$_3$Ga (or 5 layers of each sublattice), without it being exactly commensurate with the Mn$_3$Ga lattice spacing. This is completely different from the oscillations in $L1_0$ ferromagnets, which remain commensurate to the lattice. \cite{Galante2019} The oscillations we observe here have identical periods and are approximately in anti-phase (staggered) on the two magnetic sublattices. The in-plane STTk on the Mn$_\mathrm{II}$-sublattice is nearly twice as large as that on the Mn$_\mathrm{I}$ sublattice. 

In Fig. \ref{fig02} we demonstrate the dependence of the STTk oscillations, which we just described, on the dispersion of the evaluated at the Fermi-level derivative in Eq. (\ref{eq:STTk}), and their decomposition over the 4s, 4p and 3d atomic orbitals. We introduce parameter $\delta E$ to define an energy range $E_\mathrm{F}\pm \delta E/2$ around the Fermi level in which the right-hand side of Eq. (\ref{eq:kdecomp}) is averaged. It can be seen that, allowing more energy channels in the average for the spin-density's derivative with respect to bias, leads to a faster decay of the STT into the Mn$_3$Ga, without affecting much its amplitude close to the interface. Increasing $\delta E$ also smoothens the sharp features of the STTk at the interface, which are due to resonant tunneling between interface states and tend to be suppressed with larger MgO barrier thickness (see Fig. \ref{fig04}). The effect is similar for both sublattices. A small difference between the sublattices appears in the atomic orbital decomposition [Fig. \ref{fig02}(c,d)], where the relative contribution of the 4s and 4p orbitals is larger for the Mn$_\mathrm{I}$ sublattice. However, in both cases the 3dz$^2$ character dominates the calculated atomically-resolved STTk.  

As long as the adjacent Mn atoms from different sublattices have opposite sign spins, as they do, such staggered relation of the sublattice STTk is expected to result in a net STTk on the net ferrimagnetic moment, dominated by the Mn$_\mathrm{II}$ sublattice. As the atomically-resolved in-plane STTk becomes close to zero in a few MLs and changes sign, there will also be torques acting against the ferromagnetic exchange interaction in each sublattice. Torques against the AFM coupling are more subtle because of the apparently stable long-range anti-phase relation between in-plane STTk in the two sublattices, hence the two sublattices' spins are expected to rotate in the same direction. The direction of the net torque and its magnitude is thus expected to depend on the number of MLs of Mn$_3$Ga in the stack. We will demonstrate this in Section \ref{sect:last} (see Fig. \ref{fig15}). Before we focus on the electronic structure mechanism giving rise to this long-range oscillation of the in-plane STTk, we will investigate its dependence on the structural parameters of the Mn$_3$Ga analyser layer and its iterface to MgO. 

\myFig{1}{1}{true}{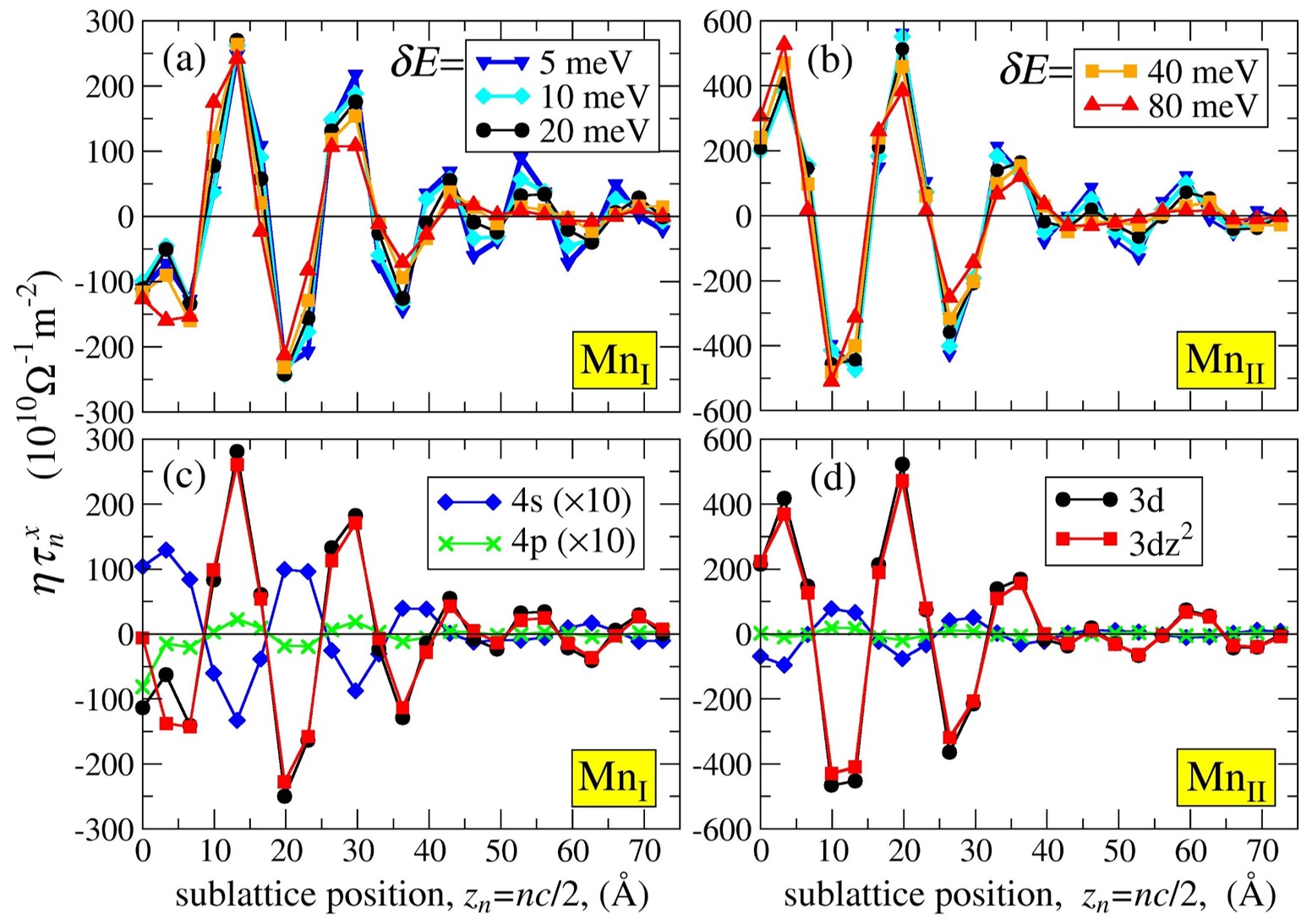}{(Color online) (a,b) In-plane STTk calculated as per Eq. (\ref{eq:STTk}), but averaging the Fermi-level expression in Eq. (\ref{eq:kdecomp}) over different finite energy ranges $E_\mathrm{F}\pm \delta E/2$ around the Fermi level and (c,d) the representative case for $\delta E=20$\,meV decomposed over orbital character, for the two Mn sublattices, as indicated in the panels. Note that 4$s$ and 4$p$ components are scaled by a factor of 10.}{fig02}

\section{Effect of the M\lowercase{n}$_3$G\lowercase{a}-M\lowercase{g}O interface and the M\lowercase{g}O barrier thickness} \label{sect:barrier}

\myFig{1}{1}{true}{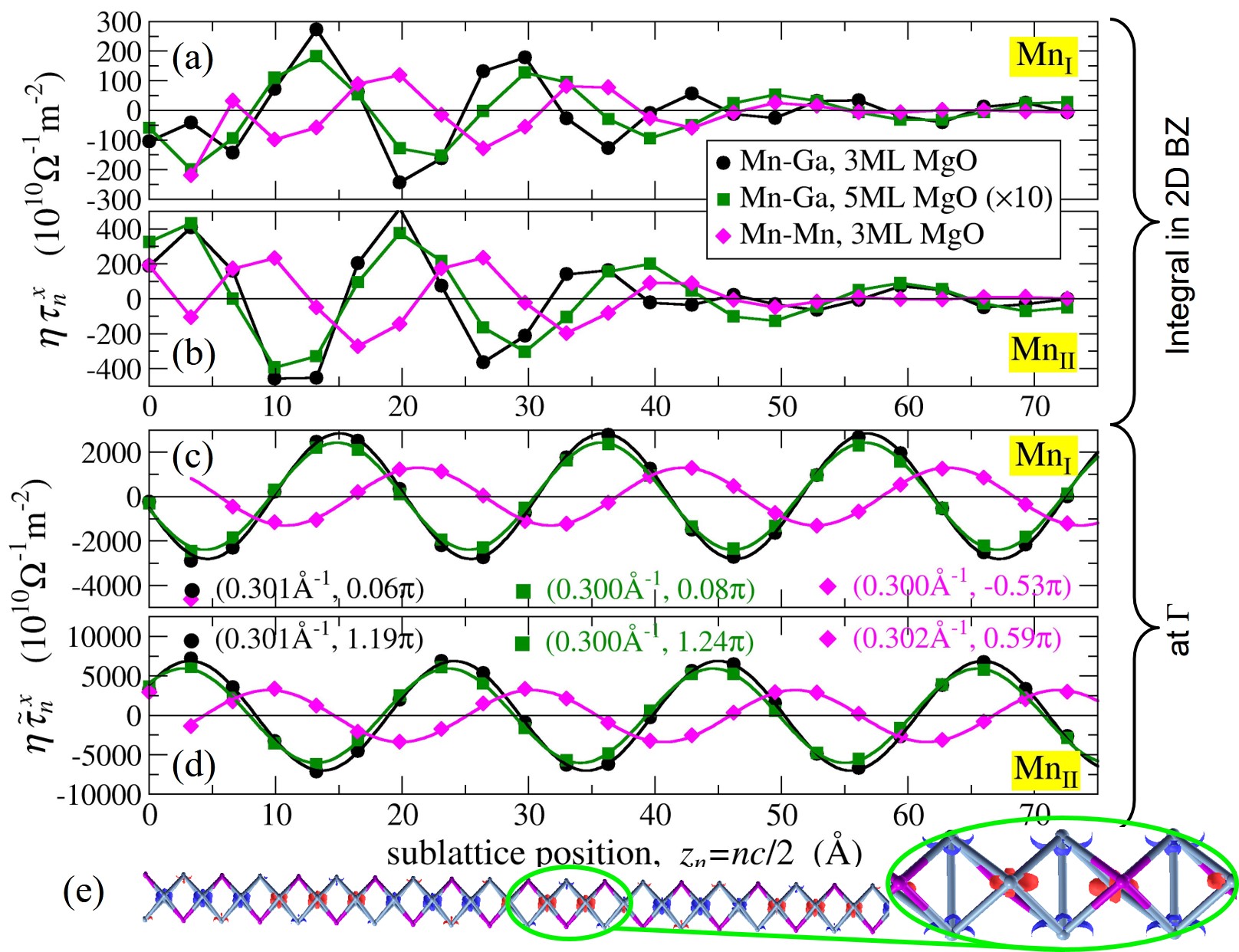}{(Color online) Comparison of atomically resolved in-plane STTk in Mn$_3$Ga for the two Mn sublattices and three different junction geometries: two interface terminations for the 3 MLs of MgO and the Mn-Ga terminated junction with 5 MLs of MgO barrier. [(a) and (b)] the total values of the on-site STTk, integrated over the transverse 2D BZ on a 80$\times$80 $k$-point mesh (the straight lines are only guide to the eye); [(c) and (d)] STTk component ($\tilde{\tau}$) calculated at the $\Gamma$ point only. Pairs of values in brackets correspond to the fitting parameters (wavenumber, phase) for the sine-wave curves shown in (c, d). In (e) is a schematic of the Mn$_3$Ga part of the junction with the $\tilde{\tau}$ iso-surfaces depicted [red (blue) are for a positive (negative)
iso-value of the STTk density]. }{fig03}

 We find that the long-range in-plane torkance oscillation is robust and is not limited to the geometry of the representative stack in Fig.~\ref{fig01}. Increasing the barrier thickness by 2 MgO MLs (from 3 to 5 MLs) results in a decrease of the in-plane STTk by an order of magnitude (see Fig.~\ref{fig03}). However, there is barely any effect on the phase of the oscillation or its carrier wavenumber in the first one or two periods. Arguably, there is a somewhat longer-ranged decay of the STTk oscillation in the case of a thicker barrier but we do not aim to quantify this effect, likely related to the enhanced directional and spin filtering. The decay of the total STTk, arising from the integration in the 2D BZ of Eq. (\ref{eq:kdecomp}), is a result of the self-cancellation from the superposition of sinewave-like oscillations from a wide range of different $\vcr{k}$ channels \cite{Stiles2002}. If we only consider the STTk at the $\Gamma$-point, i.e. $\tilde{\vcrg{\tau}}\equiv\tau\SqB{\vcr{k}_\perp=\RnB{0,0}}$, we find a perfect sine wave spatial oscillation [Fig.~\ref{fig03}(c,d)]. Furthermore, at the $\Gamma$ point the oscillation frequency and phase show no dependence on the thickness of the barrier. The fitted wavenumber in both cases is $\vcrg{\kappa}=\RnB{0.301\pm 0.001}$\,\AA$^{-1}$. Note that the values of the STTk at the $\Gamma$ point are much higher than the integral values, which are normalised to the 2D BZ area. As we will see later (e.g. in Fig.\ref{fig08}) this is because the $\Gamma$ point has the dominant contribution to the STTk, as well as to the transmission in the junction. 

In the case of Mn$_\mathrm{II}$-Mn$_\mathrm{II}$ interface termination (realised by removing the first Mn$_\mathrm{I}$-Ga layer and restoring the same interface spacing to MgO), we find a very similar long-range oscillation (see Fig. \ref{fig03}a,b, where this is compared to the reference geometry case). The termination appears to affect significantly the STTk in the first layers from the MgO -- the interface effects are stronger in the Mn$_\mathrm{II}$-Mn$_\mathrm{II}$ termination case. Deeper into the Mn$_3$Ga layer, the difference amounts mainly to a phase shift of the oscillation. The period appears very similar, but the amplitude is somewhat reduced compared to that of the Mn$_\mathrm{I}$-Ga termination case and hence the net torque is also expected to be reduced in the Mn$_\mathrm{II}$ terminated junction (see also Fig. \ref{fig15}). Again, we do not aim for a quantitative analysis of the decaying total atomically-resolved STTk. We find that the wavenumber of the oscillation of the STTk at the $\Gamma$-point is not affected by the interface composition or the thickness of the barrier [Fig.~\ref{fig03}(c,d)]. For completeness we present also the real-space density distribution of the $\Gamma$-point STTk in Mn$_3$Ga [Fig.~\ref{fig03}(e)] in terms of iso-surfaces and note that these show a $\mathrm{dz}^2$-orbital-like angular dependence, in agreement with the results in Fig. \ref{fig02}(c,d). The stack geometry is, in general, unrelaxed. As mentioned, the LSDA relaxation, under the lateral constraints of bcc Fe, results in a significant uni-axial compressive strain to the tetragonal Mn$_3$Ga and the interface to MgO splits apart. However, in order to obtain some guidance for the atomic interface reconstruction from this level of DFT, we have performed a number of relaxations of a slab of Mn$_3$Ga/MgO, subject to constant-volume constraints. In this way we have arrived at an interface distance between the Mn$_\mathrm{I}$-Ga plane and the first MgO plane of $d=2.215$~\AA (an average of the Mn-O and Ga-O bond lengths), minimizing total energy, and this is used in our representative junction geometry throughout the paper. This value has been further supported by observations described in Section \ref{sect:intro}. 

In order to rule out possible artefacts related to the chosen inter-layer distance, we have also calculated the atomically-resolved in-plane STTk for two smaller values of $d$, reduced by 0.2~\AA~ and 0.4~\AA~ (see Fig. \ref{fig04}). Without analysing quantitatively the total STTk as a function of the distance from the interface, we see that the long-range oscillation is present for smaller $d$ values too and, although there is phase shift and change in amplitude, the period of the oscillation appears very similar [Fig.~\ref{fig04}(a,b)]. Similarly to our representative case, the oscillations in the two sublattices also remain staggered. The $\Gamma$-point analysis [Fig.~\ref{fig04}(c,d)] confirms a monotonic phase shift as $d$ is decreased, and a small, non-monotonic, change in the amplitude. Both sublattices are affected similarly by the change in $d$. The fitted wavenumber ($2\pi/\lambda$) in all cases, however, remains the same: $\kappa=\RnB{0.300\pm0.001}$~\AA$^{-1}$, showing no sensitivity to interface scattering properties and hinting to the likelihood of it to be a manifestation of only bulk electronic structure properties of Mn$_3$Ga.  

\myFig{1}{1.1}{true}{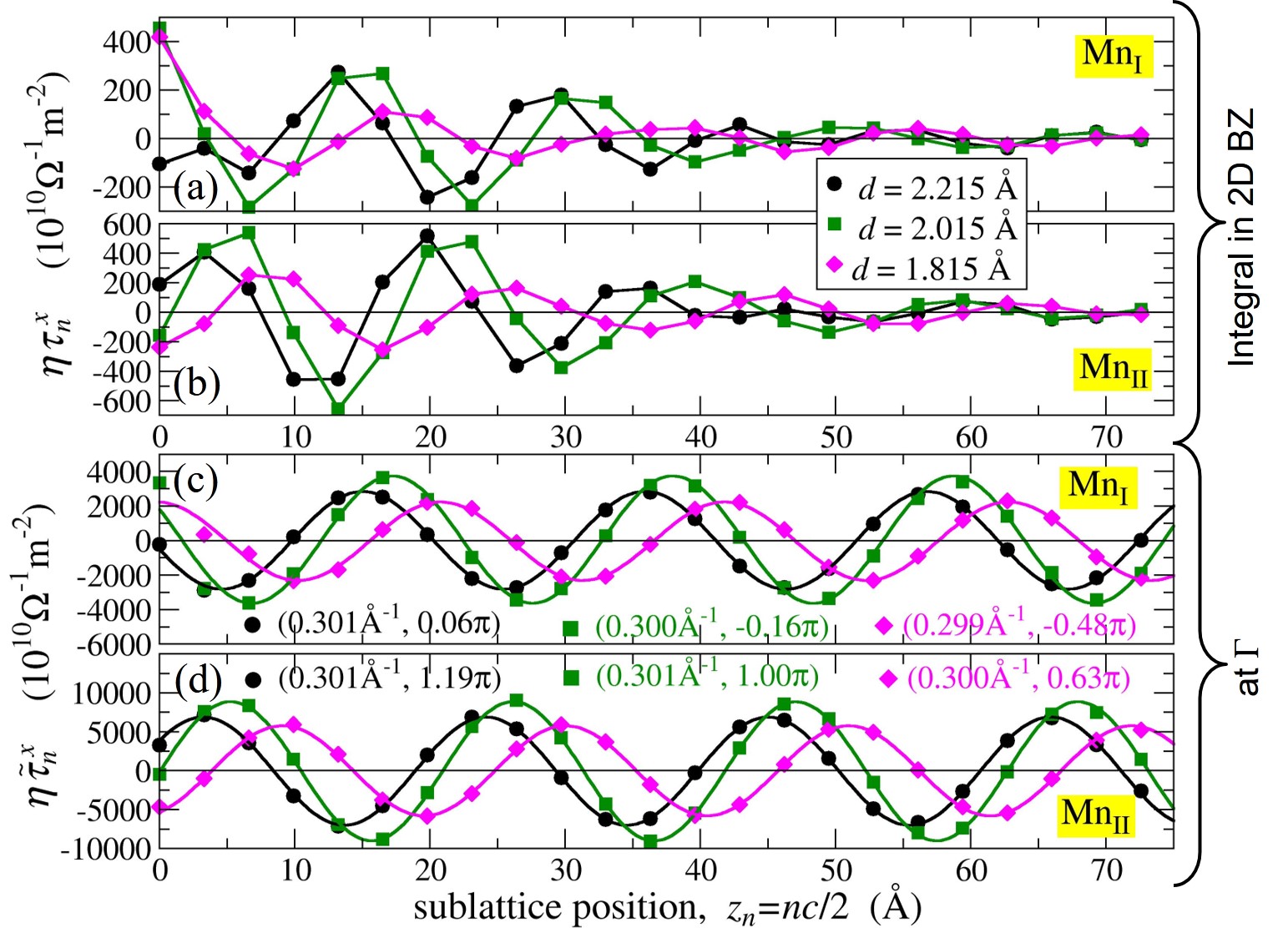}{(Color online) Comparison of atomically resolved STTk in Mn$_3$Ga for the two Mn sublattices and three different values of the Mn$_\mathrm{I}$-Ga/MgO interfacial distance,  $d=2.215$\,\AA\, (reference case), 2.015\,\AA\, and 1.185\,\AA, as indicated.}{fig04}

\section{Effect of the structural properties of the
M\lowercase{n}$_3$G\lowercase{a} lead} \label{sect:lattice}

As the geometry of our multi-layered stack is constrained laterally to the lattice constant of bcc Fe, $a=b=\sqrt{2}a_\mathrm{Fe}=4.05$~\AA, and the Fe/MgO side of the junction is fixed to established structures (see Section \ref{sect:intro}), we have taken the approach to probe the few remaining free longitudinal parameters of our geometry and have started by investigating the effect of the inter-layer spacing $d$ in the previous Section \ref{sect:barrier}. Here we explore two new values for the tetragonal lattice constant $c$ of Mn$_3$Ga, namely a smaller $c=6.4$~\AA~and a larger $c=6.8$~\AA. We then compare the calculated in-plane STTk with the original reference choice of $c=6.6$~\AA~(see Fig.~\ref{fig05}). 

We find that the long-range oscillation of the total in-plane STTk is also present for the other $c$-values and its amplitude is practically unaffected by $c$. The phase of the long-range STTk oscillation and its period, however, depend on $c$ in a monotonic way. The oscillations are again staggered between the two sublattices for all $c$-values. It is interesting, once more, to compare the in-plane STTk at the $\Gamma$ point (Fig.~\ref{fig05}c,d). There we can quantify the phase shift, which is approximately linear with $c$ for both sublattices. The amplitude is affected differently by $c$ in the two sublattices, namely there is a significant (approximately linear) increase of amplitude with $c$ for the Mn$_\mathrm{I}$ sublattice with a factor of 2.4 between $c=6.4$~\AA~and $c=6.8$~\AA, while we find a much smaller and non-monotonic variation of about 7\% for the Mn$_\mathrm{II}$ sublattice. This disbalance in sensitivity towards the inter-spin distances (and bond angles), in favor of Mn$_\mathrm{I}$ (the sub-lattice with the lower symmetry of the local environment), has been also evidenced experimentally, for example in the sensitivity of the sub-lattice moments on temperature in the MnGaRu system\cite{Betto2015}. There, the Mn sub-lattice lacking inversion symmetry (4c) has a much stronger temperature dependence, when compared to the inversion symmetric (4a) position. The sensitivity of the Mn exchange integrals on bond-lengths and bond-angles is well-established for metallic and dielectric systems alike. Here we demonstrate that the same sensitivity is propagated to the scattering properties of the electrons at the Fermi level, in particular, but not limited to, the non-directionally-averaged torkance at the $\Gamma$-point [see Fig. \ref{fig05}(c,d)].

\myFig{1}{1}{true}{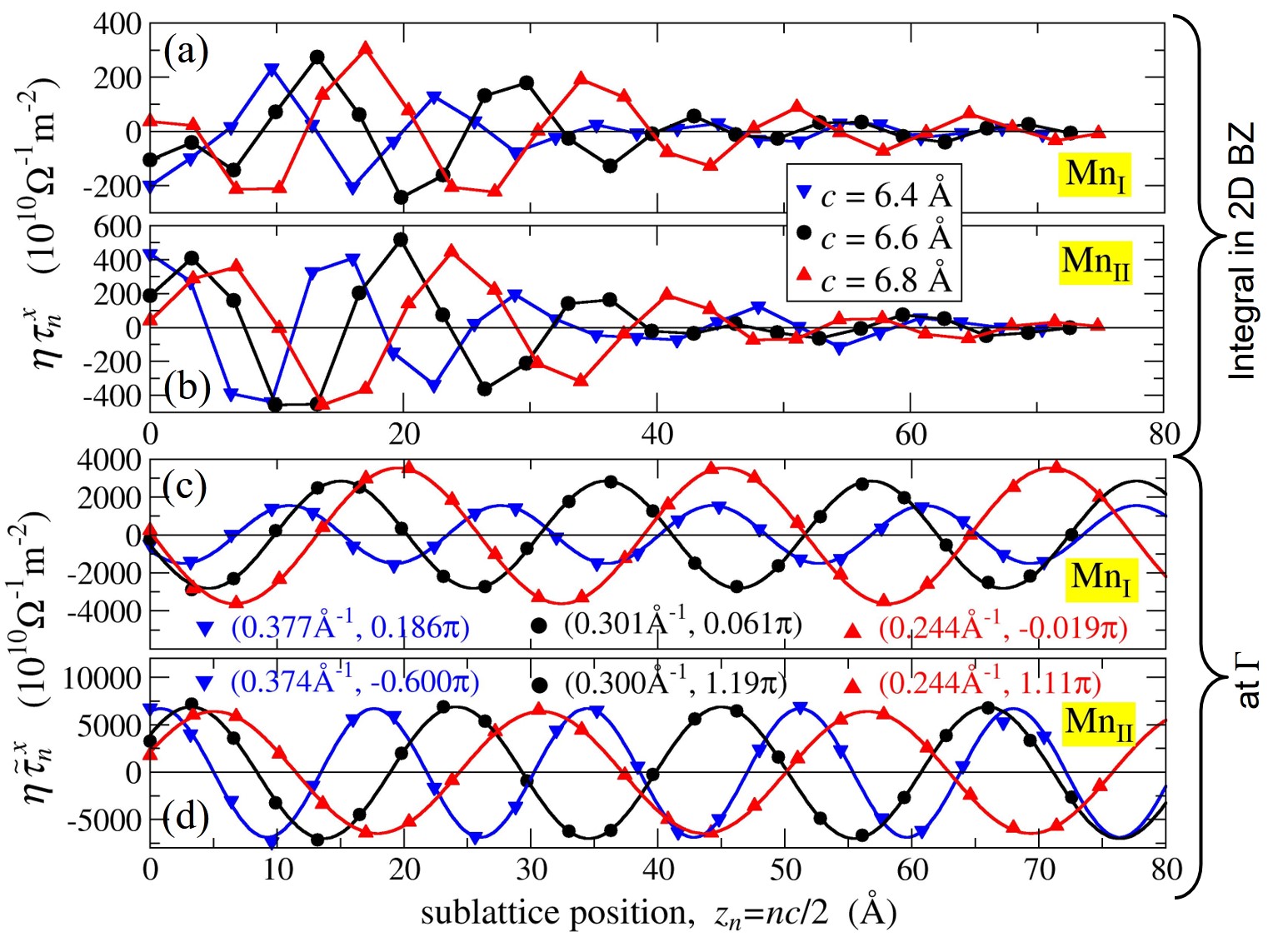}{(Color online) Effect of the longitudinal lattice constant $c$ of the tetragonal  Mn$_3$Ga. Compared to our representative lattice constant ($c=6.6$~\AA~0) are a smaller and a larger $c$-value, i.e. $c=6.4$\,\AA~(blue downward pointing triangles) and $c=6.8$\,\AA~(red upward triangles). Figure structure is analagous to Figs.~\ref{fig03} and \ref{fig04}.}{fig05}

The large variation with $c$ of the STTk amplitude in the Mn$_\mathrm{I}$ sublattice is not preserved in the total in-plane STTk, which implies different contributions from the 2D BZ -- we will investigate that later. The $\Gamma$-point in-plane STTk for all $c$ values, however, clearly shows a sine-wave oscillation with wavenumber $\kappa$ monotonically varying with $c$ [see in Fig. \ref{fig05}(c,d) insets with fitted parameters]. 
In order to understand the $\Gamma$-point in-plane STTk oscillation in the 7.6\,nm-thick Mn$_3$Ga layer in the SR of our junction, we turn to the bulk properties of tetragonal Mn$_3$Ga with unit cell as the one used for the open-boundary electrode. In Fig. \ref{fig06} we show the band structures near the Fermi level for the two spin species of Mn$_3$Ga with $a=b=\sqrt{2}a_\mathrm{Fe}=4.053$\,\AA~and the three investigated values of $c$ (6.4\,\AA, the central case of 6.6\,\AA and 6.8\,\AA) as well as that of bcc Fe with $a_\mathrm{Fe}=2.866$\,\AA~in the direction of the transport in the stack ($z$-direction, corresponding to our $\Gamma$-point transport). Highlighted are the $\Delta_1$ and $\Delta_5$-symmetry bands for spin-up $\RnB{\uparrow}$ and spin-down $\RnB{\downarrow}$ carriers. The latter two bands comprise for the leading evanescent states in the MgO barrier while the $\Delta_1$ states decay more slowly than $\Delta_5$ \cite{Butler2008} -- the TMR effect in Fe/MgO/Fe(100) is largely due to the fact that there is no matching $\Delta_1$ symmetry band at the Fermi level for the minority spins in bcc Fe [see Fig.\ref{fig06}(d)], i.e. the well-established spin-filtering effect \cite{Butler2008}.

\myFig{1}{1}{true}{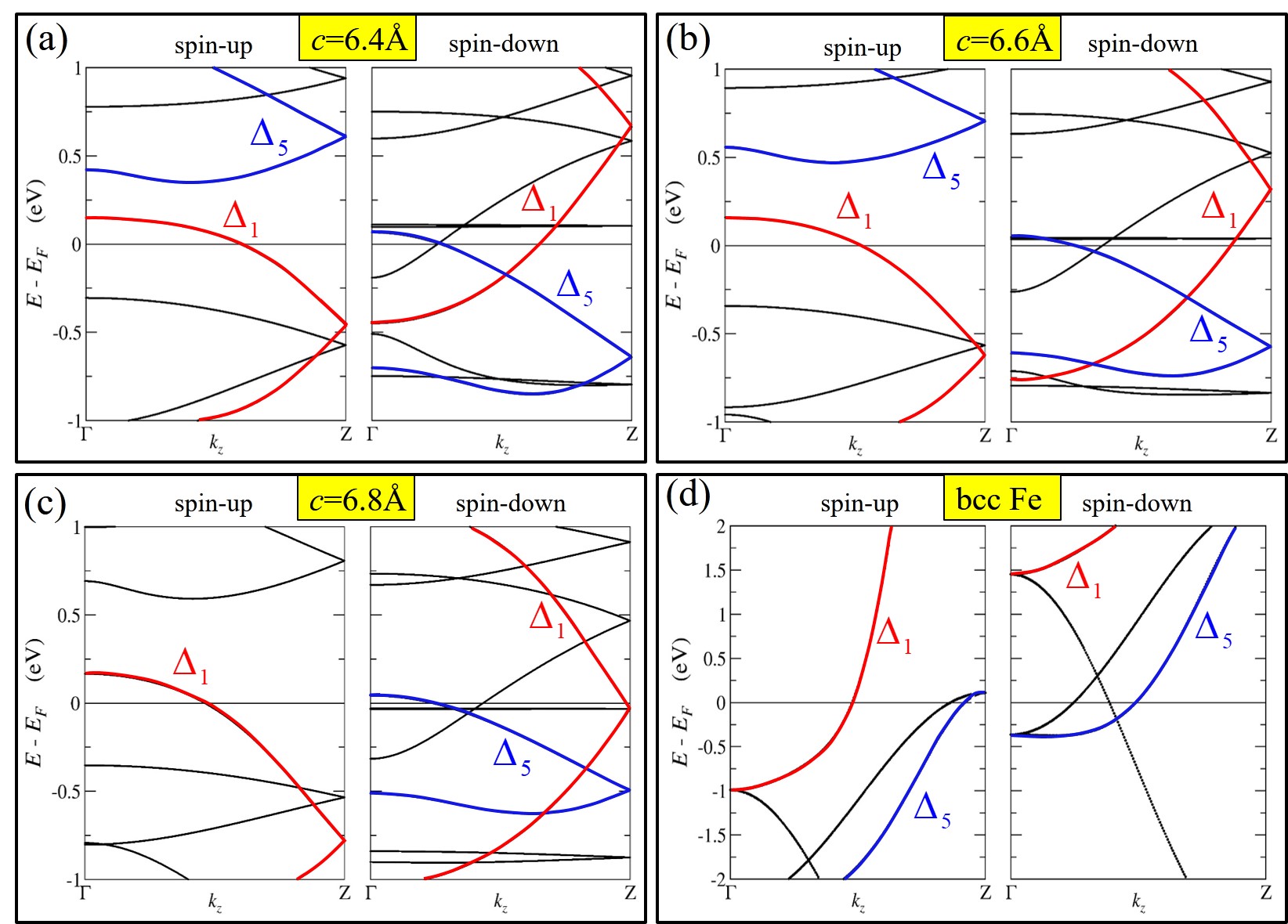}{(Color online) Band structure ($\Gamma-Z$) in the vicinity of $E_\mathrm{F}$ for three different values of the tetragonal lattice constant of bulk Mn$_3$Ga (a) $c=6.4$ \AA, (a) $c=6.6$ \AA~and (c) $c=6.8$ \AA, with $a=b=\sqrt{2}a_\mathrm{Fe}=4.053$\,\AA. The spins on the Mn atoms are aligned as in Fig. \ref{fig01}(b) and spin-up/down are defined with respect on the z-axis. Panel (d) shows the corresponding band structure of bcc Fe with $a_\mathrm{Fe}=2.866$\,\AA. Marked in thicker red and blue curves are the $\Delta_1$ and $\Delta_5$ symmetry bands, respectively. For the bcc Fe case spin-up/down correspond to majority/minority spin species.}{fig06}

It is evident from Figs. \ref{fig06} (a), (b), and (c), that the value of $c$ has little effect on the spin-up band structure around $E_\mathrm{F}$ in the $z$-direction, namely for all cases considered, there is always a single $\Delta_1$-symmetry band crossing the Fermi level. In all spin-down band structures we find both a $\Delta_1$ and a $\Delta_5$ band crossing the $E_\mathrm{F}$. This makes Mn$_3$Ga different from bcc Fe, where only a $\Delta_5$ minority-spin band crosses the Fermi level [Fig. \ref{fig06}(d)]. The lack of a minority-spin $\Delta_1$ band effectively underpins the very large theoretical values of TMR in Fe/MgO/Fe(100) junctions at low bias\cite{Butler2001}. As for Mn$_3$Ga we also find $\Delta_1$ bands for both spin-up and spin-down and we expect these to dominate the transport. Indeed, evidently also from Fig. \ref{fig03}(e), the spatial distribution of the calculated $\Gamma$-point in-plane STTk in Mn$_3$Ga is of $\Delta_1$ symmetry. 

\begin{table}[!htbp]
\centering
\begin{tabular}{ *8{|c}| }
\multicolumn{1}{c}{} & \multicolumn{3}{|c|}{From transport} & \multicolumn{4}{c|}{From bulk} \\
\hline
\rule{0pt}{3ex}$c$ & $\overline{S^z_\mathrm{Mn_I}}$ & $\overline{S^z_\mathrm{Mn_{II}}}$ & $\kappa (\tau^x$) & $S^z_\mathrm{u}$ & $k_\mathrm{F}( \Delta_1^{\uparrow})$ & $k_\mathrm{F}( \Delta_1^{\downarrow})$ & $k_\mathrm{F}( \Delta_5^{\downarrow})$ \\
(\AA) & ($\mu_\mathrm{B}$) & ($\mu_\mathrm{B}$) & (\AA$^{-1}$) & ($\mu_\mathrm{B}$) & (\AA$^{-1}$) & (\AA$^{-1}$) & (\AA$^{-1}$) \\
\hline
\rule{0pt}{3ex}6.4 & 3.33 & -2.20 & 0.376 & -1.98 & 0.291 & 0.316 & 0.128 \\
6.6 & 3.44 & -2.39 & 0.301 & -2.51 & 0.249 & 0.403 & 0.109 \\
6.8 & 3.51 & -2.56 & 0.244 & -3.12 & 0.212 & 0.468 & 0.111 \\
\hline
\end{tabular}
\caption{Properties of Mn$_3$Ga extracted from 0 V transport and from bulk ground state calculations for three values of the tetragonal lattice constant $c$. The average local spins (Mulliken population) , $\overline{S^z_\mathrm{Mn_{I,II}}}$, on the two Mn sublattices in the SR show a good agreement with the net spin $S^z_\mathrm{u}$ of the tetragonal unit cell obtained from a LSDA calculation of bulk Mn$_3$Ga, i.e. $S^z_\mathrm{u}\simeq 2\RnB{\overline{S^z_\mathrm{Mn_{I}}}+2\overline{S^z_\mathrm{Mn_{II}}}}$. The Fermi wavevectors $k_\mathrm{F}( \Delta_{1,5}^{\uparrow,\downarrow})$ in the $\Gamma$-Z direction of bulk Mn$_3$Ga, and the fitted spacial precession wavenumber $\kappa$ of the in-plane STTk at $\Gamma$ [in Fig.\ref{fig05}(c,d)], are further compared graphically in Fig. \ref{fig07}.}
\label{table1}
\end{table}

The values of the Fermi wavevector for the $\Delta_1^{\uparrow}$, $\Delta_1^{\downarrow}$ and $\Delta_5^{\downarrow}$ bands for the different lattice constants $c$ are listed in Table \ref{table1}. Also given in the table are the fitted wavenumbers from the in-plane STTk spatial precession in Mn$_3$Ga in Fig. \ref{fig05}, as well as the average local moment (from Mulliken population analysis) on the two Mn sublattices in the SR. As we can expect, the increase of the lattice constant corresponds to an increase of the local moments on Mn. The Mn$_\mathrm{II}$ sublattice is more affected, namely for the increase of 6.3\% in $c$ between 6.4 and 6.8\,\AA, we calculate, for the averaged over all atoms in the SR z-components of the Mn spins $\overline{S^z}$, an increase of 5.4\% for $\overline{S^z_\mathrm{Mn_{I}}}$ and 16.4\% for $\overline{S^z_\mathrm{Mn_{II}}}$. At the same time there is about 36\% decrease in $\kappa$, the fitted wavenumber of the in-plane STTk at the $\Gamma$ point, showing a significant sensitivity on $c$. The basic scattering theory considerations in Ref. [\onlinecite{Stiles2002}] offer an insight into the spatial precession of the STT -- for a single channel it is expected to exhibit an oscillatory behaviour of the form $\sim \exp\SqB{i\RnB{k_\uparrow - k_\downarrow}z}$. Their analytical result for the free-electron model $k$-space integration gives an oscillatory decay governed by majority-minority Fermi wavevector difference. In Fig. \ref{fig07} we compare $\kappa$ as a function of $c$ to differences of Fermi wavevectors between the only available in the $z$-direction spin-up band and the two most significant for the MgO tunneling spin-down bands, namely the $\Delta_1$ and the much more attenuated $\Delta_5$.               

\myFig{1}{1}{true}{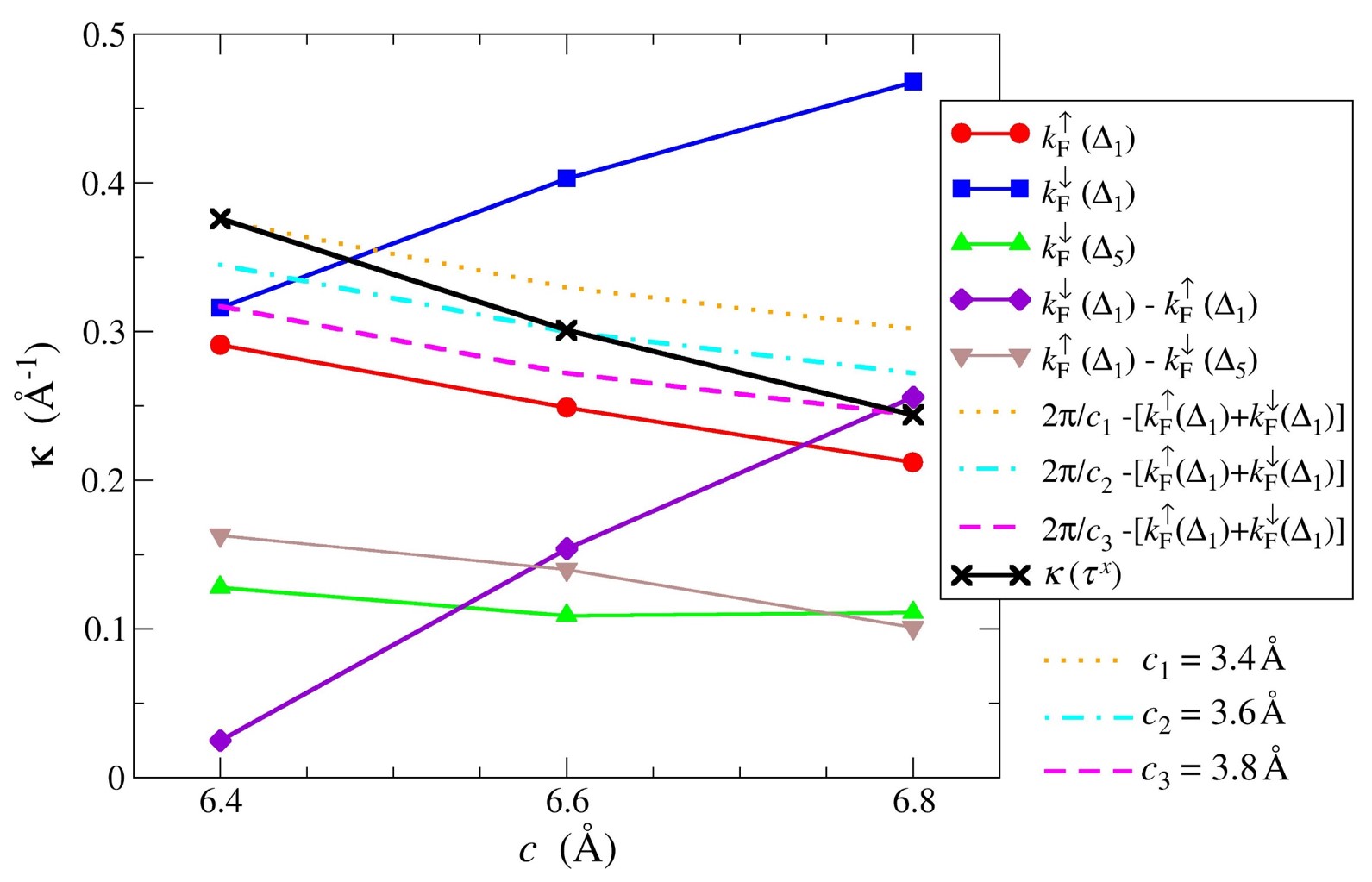}{(Color online) Calculated Fermi wavevectors for bulk Mn$_3$Ga as a function of the $c$ lattice constant as per Table \ref{table1} and a few possible differences of spin-up and spin-down Fermi wavevectors (lines are guide to the eye). Dashed lines represent $2\pi/c - k_\mathrm{F}\RnB{\Delta_1^\uparrow} - k_\mathrm{F}\RnB{\Delta_1^\downarrow}$ for all three values of $c$. The black "x" symbols correspond to the fitted wavenumbers $\kappa(c)$ of the $\Gamma$-point in-plane STTk oscillation from Fig. \ref{fig05}c,d. }{fig07}

It is evident from Fig. \ref{fig07} that using the $\Delta_5^\downarrow$ band Fermi wavevector, i.e. $k_\mathrm{F}(\Delta_1^\uparrow)-k_\mathrm{F}(\Delta_5^\downarrow)$, results in a much smaller spatial oscillation frequency than the fitted $\kappa$ from Fig. \ref{fig04}(c,d). At the same time, taking directly $k_\mathrm{F}(\Delta_1^\downarrow)-k_\mathrm{F}(\Delta_1^\uparrow)$ does not even reproduce the correct sign of the slope. We notice that the group velocities at the Fermi level along $z$ have opposite signs for the spin-up and spin-down (Fig. \ref{fig06}) in this part ($k>0$) of the first BZ for two of the $c$ values. In order to make sure we consider carriers travelling in the same direction we take the negative image of the already calculated $k_\mathrm{F}(\Delta_1^\downarrow)>0$ (which currently describes right-going states in the junction for $c=6.4$\,\AA~ and $c=6.6$\,\AA, and is practically equal to the BZ boundary wavevector for $c=6.8$\,\AA), i.e. we substitute $k_\mathrm{F}(\Delta_1^\downarrow)\rightarrow-k_\mathrm{F}(\Delta_1^\downarrow)$, and add a shift by the reciprocal lattice vector $2\pi/c$. Hence we calculate $2\pi/c - \SqB{k_\mathrm{F}(\Delta_1^\uparrow)+k_\mathrm{F}(\Delta_1^\downarrow)}$, which is plotted for all three $c$ values in Fig. \ref{fig07} and shows a remarkable agreement with the fitted spatial frequency $\kappa$. It is clear that at $\Gamma$ the spatial precession of the STTk is driven by the mismatch of the majority and minority Fermi wavevectors, $k_\mathrm{F}^\uparrow-k_\mathrm{F}^\downarrow$, of the $\Delta_1$-symmetry band (which in this case results from hybridisation between $s$ and $dz^2$ orbitals) in Mn$_3$Ga, as described by the free electron model in Ref. \onlinecite{Stiles2002}. This result corroborates with the observed $dz^2$-orbital-like character of the spatial distribution of the STTk at $\Gamma$, as shown in Fig. \ref{fig03}e. The matching wavevectors at $\Gamma$ provide sufficient evidence for the nature of the spatial oscillation of the integral STTk that we observe in Mn$_3$Ga and which is a robust effect persisting for a range of lattice parameters. In the following Section \ref{sect:finbias} we will look in more detail at the decomposition of the transmission coefficients and the atomically-averaged STTk over the transverse 2D BZ. This will further elucidate the special role played by the $\Gamma$ point for the spin-transport properties of the Mn$_3$Ga/MgO/Fe(001) junctions.      

\section{Analysis of zero-bias transmission and finite-bias STT} \label{sect:finbias}

In Fig. \ref{fig08} we return to our representative structure with $c=6.6$\,\AA~ and present the decomposition of transport properties at the Fermi level energy over the 2D transverse BZ. These include: the numbers of open channels (bands crossing the Fermi level) of the two semi-infinite leads, the transmission coefficients $T_{\vcr{k}}^\sigma (E_\mathrm{F})$ for $\sigma=\uparrow,\downarrow$ and for two barrier thicknesses, and the in-plane STTk, all as functions of $\vcr{k}_\perp=(k_x,k_y)$. Within the NEGF formalism, the calculated $\vcr{k}_\perp$-dependent transmission coefficients are defined as (see e.g. Ref. [\onlinecite{Rungger2018}])
\begin{equation}
T_{\vcr{k}}^\sigma(E_\mathrm{F})=\mathrm{Tr}\left[\Gamma^\sigma_{\vcr{k},\mathrm{L}}(E_\mathrm{F}) G_{\vcr{k}}^{\sigma\dagger}(E_\mathrm{F}) \Gamma^\sigma_{\vcr{k},\mathrm{R}}(E_\mathrm{F}) G_{\vcr{k}}^\sigma(E_\mathrm{F}) \right], \label{eq:transm}
\end{equation}
where ``Tr'' denotes the matrix trace operation and $\vcr{k}\equiv \vcr{k}_\perp$ (for compactness). A persistent feature in most of the contour plots, shown in Fig. \ref{fig08}, is the dominant contribution of the $\Gamma$ point. We find, at the $\Gamma$ point, 6 open channels for both majority and minority spin in bcc Fe, while there is only one majority spin in Mn$_3$Ga (see also the band structure in Fig. \ref{fig06}) and 4 open channels for minority spin (note, $\Delta_5$ band is doubly-degenerate). The large peak in the transmission at $\Gamma$ for spin-up AP and spin-down P is due to the dominant $\Delta_1$ transmission available for majority spin both in Mn$_3$Ga and in Fe. Note, that if the two sub-lattices in Mn$_3$Ga were equivalent, the P an AP states of the junction would differ only by the spin orientation in a single interfacial ML (upon complete spin-reversal in the junction). The AFM limit elucidates the similarities in the diagonal pairs from the 2$\times$2 panels of $\vcr{k}_\perp$-resolved transmissions in Fig. \ref{fig08}(b). Increasing the MgO barrier thickness to 5 MLs accentuates the similarities between those pairs of configurations, as well as the apparent dominant contribution of the $\Gamma$ point in $\CrB{T^\uparrow_{\vcr{k},\mathrm{AP}},T^\downarrow_{\vcr{k},\mathrm{P}}}$, compared to the $\CrB{T^\uparrow_{\vcr{k},\mathrm{P}},T^\downarrow_{\vcr{k},\mathrm{AP}}}$ pair.   

\myFig{1.45}{1.45}{true}{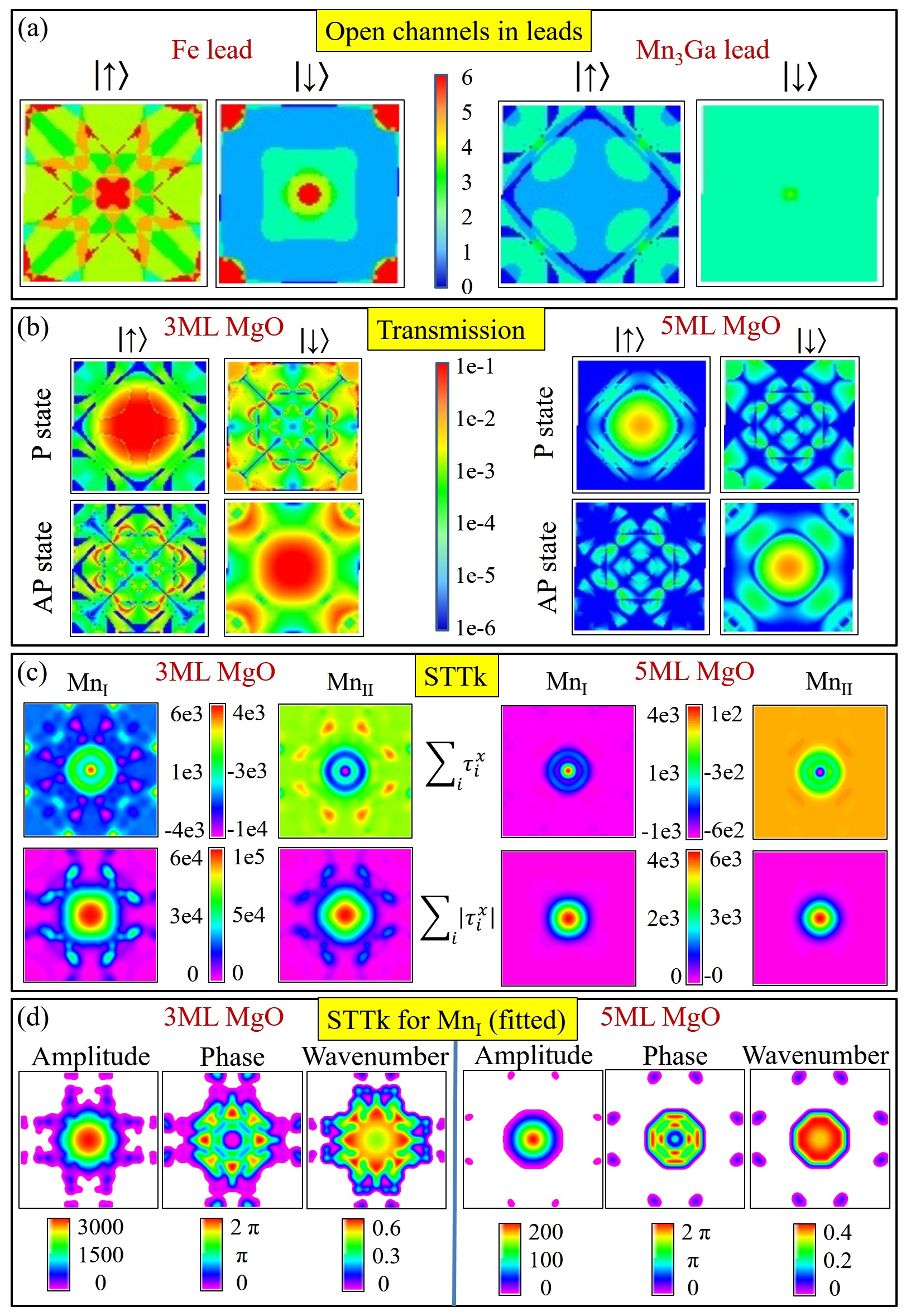}{(Color online) Contour plots of $\vcr{k}_\perp$-dependent properties in the transverse 2D BZ: (a) the open channels for the two spin species in the Fe lead and the Mn$_3$Ga electrode, respectively; (b) the transmissions for each spin species in both P and AP spin alignments; (c) two different sums of the atom-resolved in-plane STTk in the Mn$_3$Ga layer; (d) parameters of the sine-wave fits to the atom-resolved in-plane STTk in the Mn$_\mathrm{I}$ sublattice for each $\vcr{k}_\perp$ channel. Presented are two Fe/MgO/Mn$_3$Ga stacks, with 3 and 5 MLs of MgO in the left- and the right-hand side panels, respectively. Note that the hue color shade in (a,c,d) is on linear scale, but in (b) the color scale is logarithmic. See the text for details.}{fig08}

In Fig. \ref{fig08}(c) we look into analogous contour plots in the 2D BZ for the in-plane STTk in the 7.6\,nm thick layer of  Mn$_3$Ga in the SR (as per Fig. \ref{fig01}a). We first show the total in-plane STTk for the two Mn sublattices, which is defined for each $\vcr{k}_\perp$ as the sum of $\vcrg{\tau}_n (\vcr{k}_\perp)$ [from the $\vcr{k}_\perp$-decomposed version of Eq. (\ref{eq:STTk})] for all the Mn atoms in the SR from the corresponding sublattice (top panels). Then we also evaluate another quantity, $\sum \Abs{\vcrg{\tau}_n}$, for the two sublattices. The later contrast is not physically measurable, but offers additional insight about the distribution of the main contributions (as absolute values) in the 2D BZ. It is useful in comparison with the panels above, where the direct summation of atomically-resolved STTk portraits are more sensitive to the length of the SR due to the oscillatory nature of the in-plane STTk in space. The 'absolute-value contrast' also elucidates the similarities in the $\vcr{k}_\perp$-portraits of the STTk and the transmission above -- arguably it is an amalgamate of the majority and minority transmissions. We find, as expected, that the main contribution to the in-plane STTk arises from the $\Gamma$ point. This is more evident in the case of the thicker barrier (5 MLs of MgO), where the transport is even further suppressed to a small nearly-circular zone around $\Gamma$, which contributes to the STTk in Mn$_3$Ga. In this area we see that the net STTk in the 7.6\,nm Mn$_3$Ga slab changes sign in concentric rings around the $\Gamma$ point. The $\Gamma$ point contribution has a different sign for the two sublattices. Apart from the quantitative difference and some symmetry-driven shape-shift of the main contributing area around $\Gamma$, similarities between the two sublattices (even more evident in the 'absolute-value contrast'), suggest a common underlying transport mechanism for the in-plane STTk in the two sublattices. 

We further analyse the calculated $\vcr{k}_\perp$-resolved in-plane STTk in Mn$_3$Ga by performing sine-wave fit to the atomically-resolved STTk for each $\vcr{k}_\perp$-point in the 2D BZ. Because of the observed similarities of the STTk in the two sublattices, we only examine Mn$_\mathrm{I}$, and the results are shown in Fig. \ref{fig08}(d). The pattern of the fitted amplitude matches that of the 'absolute value contrast' in the panels above. Note that the white regions are cut off because of the very poor $\chi^2$ fitting parameter value (below certain threshold; in fact, at the boundary with the white region, we tend to find very abrupt apparent failure of the single sine-wave fit). Besides the clear evidence of directional filtering between the 3-ML and the 5-ML stacks, we find that the $\Gamma$ point contribution to the STTk in both cases arises at an intermediate spatial frequency and a markedly different phase compared to its surrounding area. The significant area of the 2D BZ, where wavenumber can be fitted, is indicative to the faster decay of the STTk oscillation in the 3-ML case [see Fig. \ref{fig03} (a, b)]. We would expect this decay to be clearly suppressed for thicker barriers but our calculations also show significant active 2D BZ area for STT in the 5ML case and a very small change in the decay rate\footnote{It is difficult to quantify the decay rate for this system size and the 'rectangular wave packet' fit from  Fig. \ref{fig01} does not show a difference in the dispersions on the wave number between the two barrier thicknesses.} of the in-plane STTk into Mn$_3$Ga. 

\myFig{1}{1}{true}{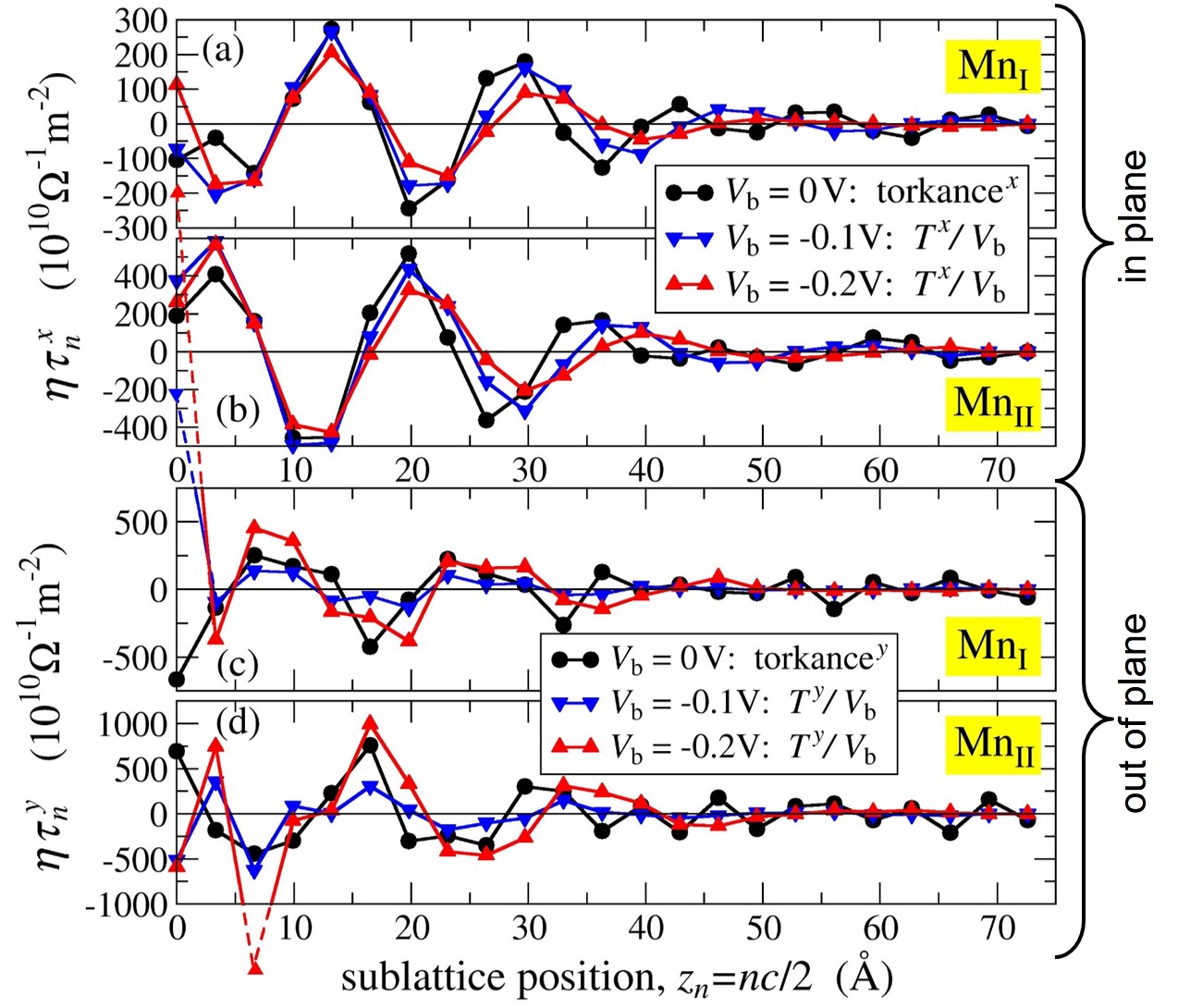}{(Color online) Comparison of linear response STTk and finite bias STT for two bias voltages $V_\mathrm{b}$=-0.1,\,-0.2\,V, scaled by their corresponding $V_\mathrm{b}$ values, as a function of the atomic position on the two Mn sublattices of the Mn$_3$Ga inside the SR. In (a,b) is the in-plane ($x$) component of the STT, while in (c,d) is the field-like ($y$) component. Straight lines between datapoints are only guide to the eyes. Note that three datapoints are shown outside their panels and connected with dashed lines. This is done to maximise resolution for the rest of the dataset.}{fig09}

To this moment we have only considered the linear response regime and now in Fig. \ref{fig09} we present the resulting STTs from self-consistent calculations at two different finite biases (in this case we have chosen $V_\mathrm{b}<0$, i.e. electrons flowing from the Mn$_3$Ga lead), scaled by their corresponding $V_\mathrm{b}$ values, in comparison to the STTk results. Our methodology for the finite-bias STT is described in Ref. [\onlinecite{Rungger2018}]. Note that the self-consistent finite bias calculations are much more challenging numerically and there is a further faster-than-linear scaling of the computational time with the bias voltage. Hence, the $V_\mathrm{b}$ we can apply is limited by the already significant size of the SR. In Fig. \ref{fig09}(c,d) we also present the out-of-plane (field-like) STT and STTk. These results demonstrate that the long-range oscillation is not an artefact of the linear response regime. It can be anticipated that opening the bias window dampens the spatial precession (in the sense of Ref. [\onlinecite{Stiles2002}]) of the STT because of the additional integration over energy (together with that over $\vcr{k}_\perp$) for the spin accumulation. Indeed, such enhanced decay is visible in all panels of Fig. \ref{fig09} and it tends to increase with the bias voltage. Even at the highest bias considered (-0.2 V), there are at least two full periods of STT oscillation visible, with similar periods to the ones observed in the linear response regime. In fact, it is clear that for the bias voltage considered, the linear response regime offers quite a good approximation for the magnitude of the STT in Mn$_3$Ga at low bias, especially close to the interface. The main effect of opening the bias window is the enhancement of the decay and, arguably, a small shift of the oscillation frequency towards larger wavelengths. Although the out-of-plane (field-like) torque is notoriously challenging to calculate accurately (requiring very high $\vcr{k}_\perp$-point sampling), we can see that despite the noise it clearly shows an oscillatory behaviour too. It appears to be offset by, roughly, a $\pi/2$ phase shift from the in-plane STT and again the two sublattices oscillate in antiphase. Interestingly, in our finite-bias calculations we find a huge out-of-plane torque on the first Mn$_\mathrm{I}$ atom at the interface. We know this site in the 3-ML junctions is affected by spin-polarised interface states and it appears that at finite bias it this is manifested in a very large field-like torque. This observation, which is not captured in the linear response regime, deserves a further investigation which goes beyond the scope of this work.      

\myFig{1.2}{1.2}{true}{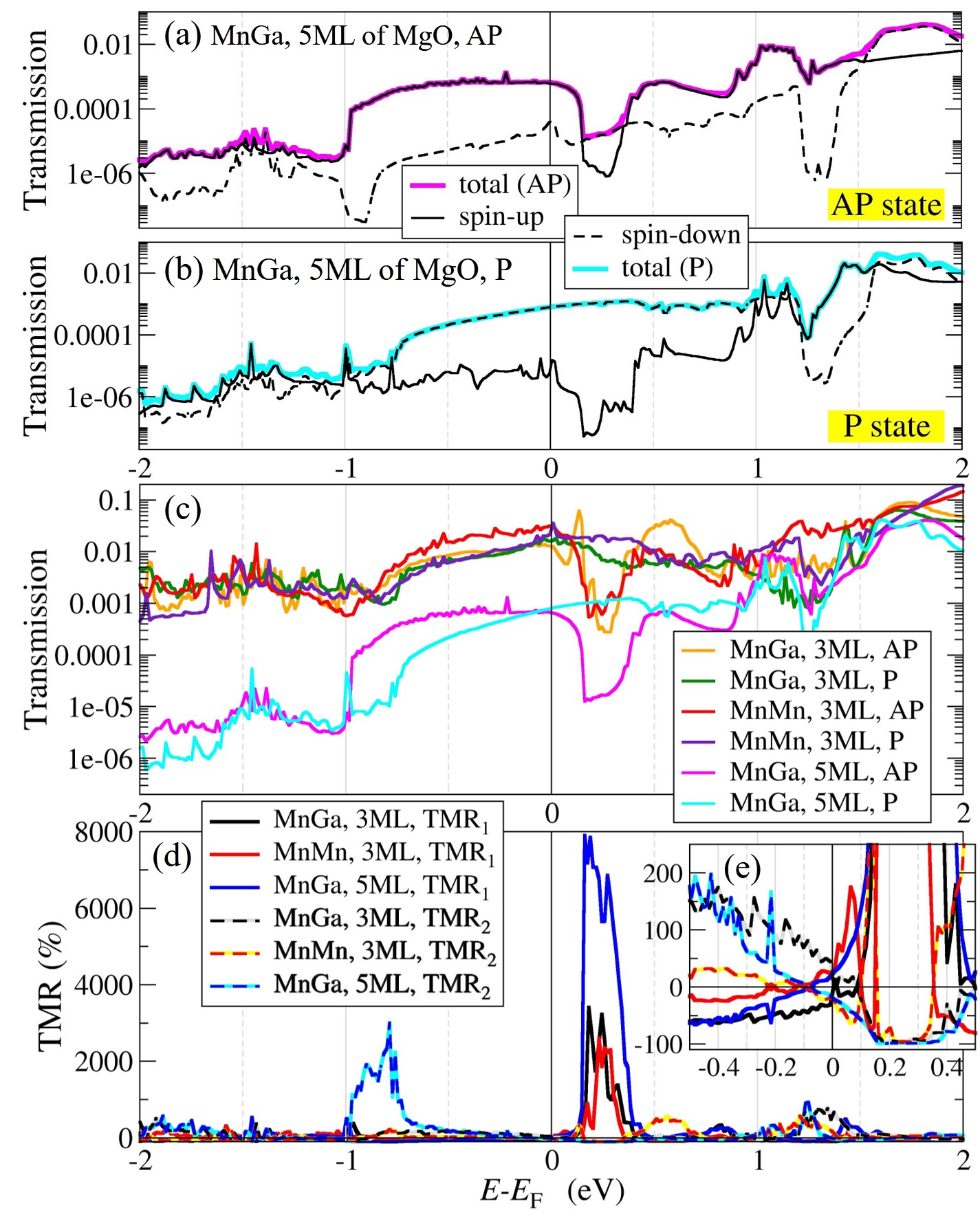}{(Color online) Energy dependence of transport properties at 0 V. In (a) and (b) are the total and the spin-dependent transmissions of the Mn-Ga terminated junction with 5 MLs of MgO in its AP and P state, respectively. In (c) and (d) are the total transmission and TMR effects [Eq. (\ref{eq:TMR})] for three different junction geometries (as indicated in the legend: the interface termination, the MgO barrier thickness and the spin state of the junctions). The inset (e) is a zoom around the Fermi-level. In (d,e) thick lines correspond to $\mathrm{TMR}_1$, while dashed lines depict $\mathrm{TMR}_2$ as defined in Eq. (\ref{eq:TMR}). }{fig10}

We have seen that the $\vcr{k}_\perp$-resolved transmission at the Fermi level shows a significant spin polarisation. In Fig. \ref{fig10} (a,b) we compare the ballistic transmission coefficients [from Eq. (\ref{eq:transm}) integrated over the 2D BZ] for the two spin species in the two, AP or P, magnetic configurations [see Fig. \ref{fig01} (f,g)] of the Mn$_\mathrm{I}$-Ga-interfaced junction with 5 MLs of MgO as function of the energy of the carriers. In the AP state Mn$_\mathrm{I}$ is aligned "up" which corresponds to the band structure in Fig. \ref{fig07}b, and also parallel to the moment of the Fe (also pointing "up"). Therefore, the conduction is dominated by the $\Delta_1^\uparrow$ band at the Fermi level. The drop in the spin-up transmission at around 0.15 eV corresponds to the $\Delta_1^\uparrow$ band edge at $\Gamma$. The other drop in the $T^\uparrow_\mathrm{AP}$ at around -1 eV is due to the other band edges of the $\Delta_1^\uparrow$ band in both Mn$_3$Ga and in bcc Fe. In comparison, the transmission of the spin-down carriers in the AP state is significantly lower around the Fermi level because it is carried by the $\Delta_5^\downarrow$ band. $T^\downarrow_\mathrm{AP}$, however, dominates in the $\Delta_1^\uparrow$ band gap between 0.15-0.4 eV and above 1.5 eV, where $\Delta_1^\downarrow$ appears. 

In the P state the spin-up transmission is similar to that of the spin-down carriers in the AP state as the spin polarisation is largely dictated by the ferromagnet (the Fe lead). The spin-down transmission (now corresponding to majority spin in bcc Fe) dominates in the P case. This is due to the availability of $\Delta_1$ bands, which co-exist in Mn$_3$Ga spin-down and in bcc Fe spin-up in a wide energy range between around -0.75 eV and 1.5 eV. The band-edges of the $\Delta_1^\uparrow$ band in Mn$_3$Ga are clearly what drives the TMR effect in these junctions. 

Based on these energy-resolved spin-polarised ballistic transmission coefficients, we consider possible definitions for the theoretical TMR effect in the Mn$_3$Ga/MgO/Fe junction at equilibrium (0V). Unlike the prototypical Fe-MgO-Fe junction\cite{Rungger2009}, here we do not have a clear choice for the less-transmissive reference state. Note that we have defined P and AP based on the net-spin alignment between the Mn$_3$Ga and the Fe lead, but they correspond to the opposite alignment of the interfacial spins in the junction [see Fig. \ref{fig01} (f) and (g)]. Hence, we consider one definition, based on the net spin alignment ($\mathrm{TMR}_1$), and the other ($\mathrm{TMR}_2$), based on the alignment of the spins at the interface (in our representative case with Mn$_\mathrm{I}$-Ga termination) and these simply correspond to swapping the P and the AP state, thus we define
\begin{eqnarray}
\mathrm{TMR}_1 &=& \RnB{T_\mathrm{P}-T_\mathrm{AP}}/T_\mathrm{AP} \nonumber    \\
\mathrm{TMR}_2 &=& \RnB{T_\mathrm{AP}-T_\mathrm{P}}/T_\mathrm{P}\,,
\label{eq:TMR}
\end{eqnarray}
where $T_\mathrm{P,AP}(E)=T^\uparrow_\mathrm{P,AP}(E)+T^\downarrow_\mathrm{P,AP}(E)$ are the total transmission coefficients in the two spin-states of the junction. These are calculated, at equilibrium (0 V) as a function of the energy of the incoming carriers, for three different junction geometries, i.e. the two possible terminations of the Mn$_3$Ga/MgO interface and an additional thickness of 5 MgO MLs for the Mn$_\mathrm{I}$-Ga termination, and their dependence on the electron energy is shown in Fig. \ref{fig10}. Note, that this is not the typical TMR effect as function of the applied bias voltage $V_\mathrm{b}$, calculated from the current-voltage characteristics (e.g. Ref. [\onlinecite{Rungger2009}]), but a quantity indicating the spin-polarisation of the zero-bias transmission in the vicinity of the Fermi level. At $E_\mathrm{F}$, we observe a small TMR effect, showing only a small variation with geometry (see the inset) for both definitions of the TMR. Both $\mathrm{TMR}_1 (E_\mathrm{F})$ and $\mathrm{TMR}_2 (E_\mathrm{F})$ exhibit a sign change between positive and negative values, or the other way around, as the geometry changes between the Mn-Ga and the Mn-Mn termination. The absolute values of the TMR close to equilibrium for all studied cases remain between 10 and 30 \%. 

We, however find a significant TMR$_1$ effect in an 'island' from about 0.15\,eV to about $0.45$\,eV above the Fermi level. These correspond to the gap in the spin-up transmission between the band edges of the $\Delta_1^\uparrow$ and $\Delta_5^\uparrow$ bands in Mn$_3$Ga [see Fig. \ref{fig06}(b)]. The effect occurs for all geometries but is especially pronounced for the case of the thicker barrier. Similarly, in the case of 5ML MgO we also find a region of increased TMR$_2$ effect between -0.7 eV and -1 eV. This is due to the drop in $T^\uparrow_\mathrm{AP}$ at around -1 V (because of the $\Delta_1^\uparrow$ band edge in Fe) and the drop in $T^\uparrow_\mathrm{AP}$ at around -0.75 eV (because of the $\Delta_1^\downarrow$ band edge in Mn$_3$Ga). Based on these observations, we can anticipate certain features in the bias dependence of the finite-bias TMR in the Fe/MgO/Mn$_3$Ga junctions. Since the feature above the $E_\mathrm{F}$ is determined mostly by the $\Delta_1^\uparrow$ band in the Mn$_3$Ga lead, we expect this to move down in energy with applied bias voltage $V_\mathrm{b}$ at a rate of $V_\mathrm{b}/2$. Thus, this would to cause a peak in the total TMR$_1$($V_\mathrm{b}$) at positive $V_\mathrm{b}$ in the range between $0.2$ to $0.6$ V. For large negative biases (possibly above 0.5 V) we expect to see a change of sign in the TMR$_1$ ($V_\mathrm{b}$) due to the shift upward of the Mn$_3$Ga $\Delta^\uparrow_1$ band-edge driven end of the second island-like feature notable in TMR$_2$($E$) below the Fermi level. More accurate SCF finite-bias TMR calculations are subject of ongoing work.       

\section{An entirely ferrimagnetic M\lowercase{n}$_3$G\lowercase{a}-based stack} \label{sect:last}

Here we consider an analogous MTJ in which the Fe lead on the right-hand side is completely substituted with Mn$_3$Ga, namely a Mn$_3$Ga/MgO/Mn$_3$Ga stack, again with nearly 8\,nm of Mn$_3$Ga as analyser on the right-hand side [as in Fig. \ref{fig01}(a)]. Note that no lateral dimensions are changed in this rearrangement and the junctions are made mirror-symmetric with respect to the middle MgO layer [we consider only two cases again with odd number of MgO MLs, i.e. 3 and 5 MLs, and the latter is visualised in Fig. \ref{fig11}(a)]. The effect of the perfect mirror symmetry in the barrier interfaces about the central MgO layer can be immediately observed in the transmissions in Fig. \ref{fig11}d, where there is no difference between the $\vcr{k}_\perp$-resolved transmissions for the two spin species in the AP state, defined by the alignment of the interfacial Mn$_\mathrm{I}$ spins [see Fig. \ref{fig11}(b,c)], or the energy-resolved $T^\uparrow_\mathrm{AP}$ and $T^\downarrow_\mathrm{AP}$ in Fig. \ref{fig13}(a,b). Furthermore, $T_\mathrm{AP}$ appears a lot like a scaled product of $T^\uparrow_\mathrm{P}$ and $T^\downarrow_\mathrm{P}$, indicating a relative independence of the two spin-channels in this system.       

\myFig{1.0}{1.0}{true}{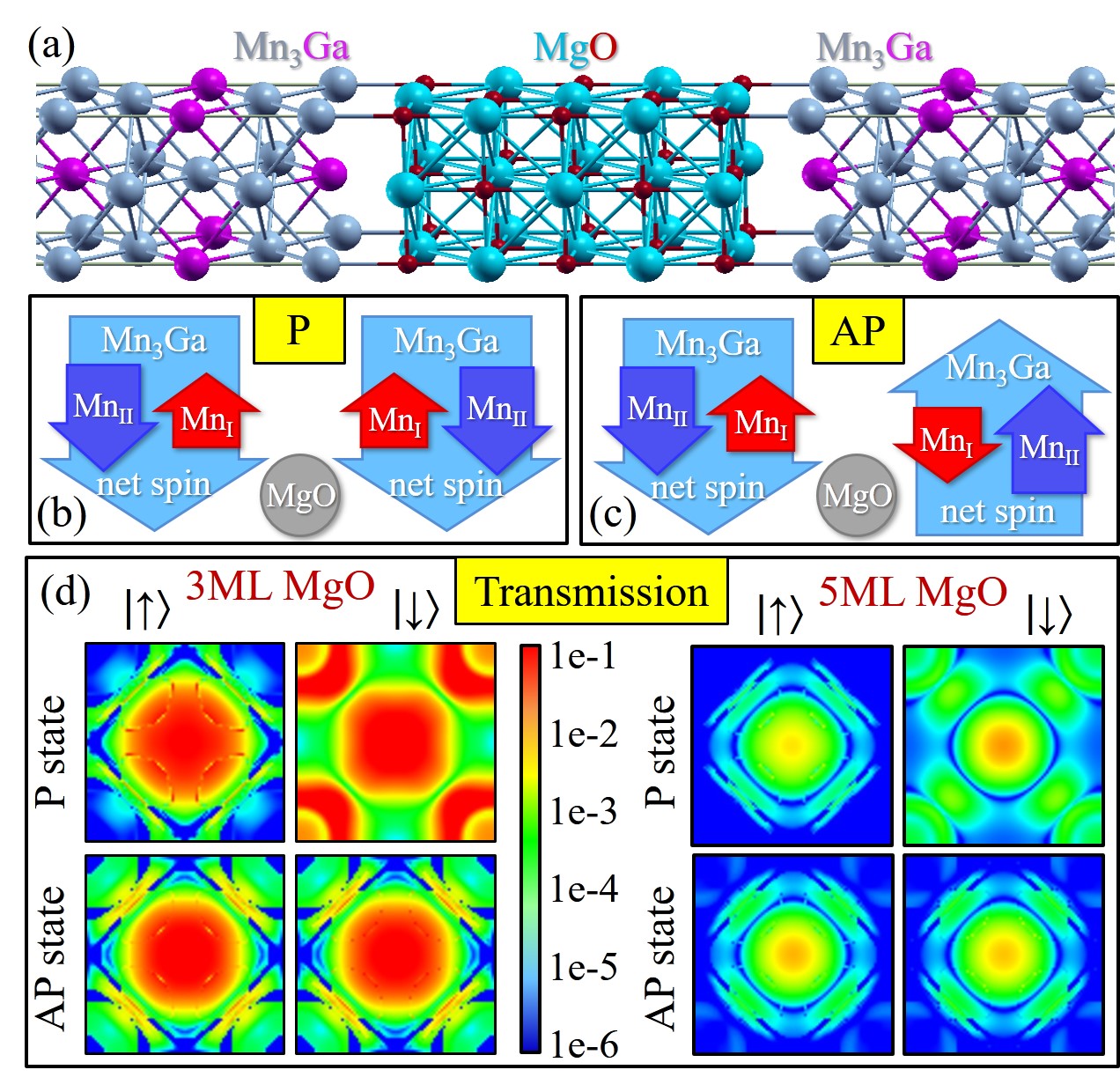}{(Color online) Schematics of: (a) the fully ferrimagnetic junction Mn$_3$Ga/MgO/Mn$_3$Ga, (b) and (c) -- the local and net spin alignments in the P and AP states, respectively. In (d) are the 2D $\vcr{k}_\perp$-portraits of the Fermi-level transmissions for the two spin species in the two spin-states of the two junctions (with 3 and 5 MLs of MgO), as indicated in the panel.}{fig11}

\myFig{1.2}{1.2}{true}{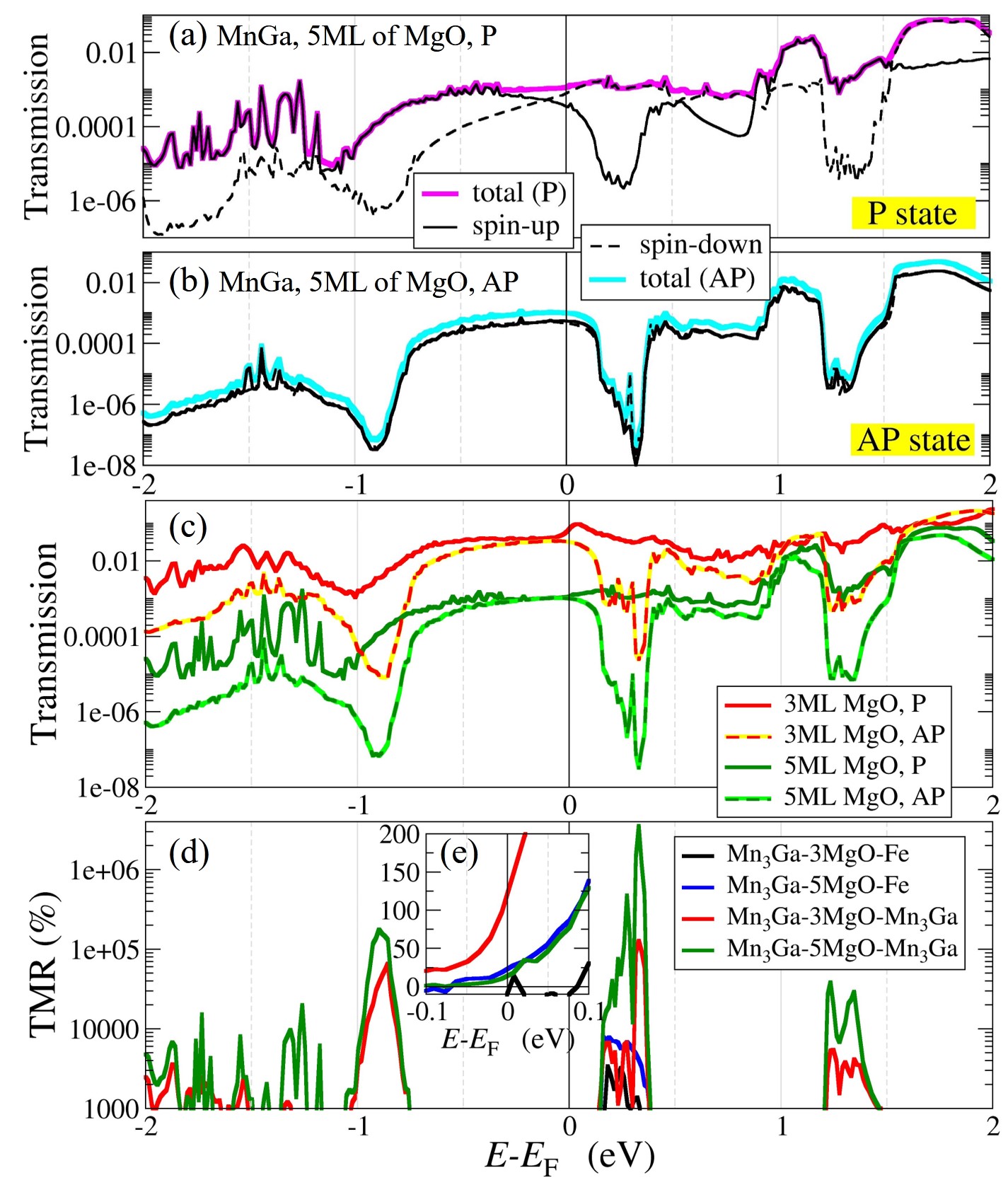}{(Color online) Energy-resolved transmission and TMR for  Mn$_3$Ga/MgO/Mn$_3$Ga stacks. In (a,b) are the spin-resolved and total transmissions for the 5ML stack in the P and the AP state, respectively. In (c) the total transmissions from above are compared to the case of the 3ML stack. In (d) is a comparison of the calculated TMR effect for the two all-ferrimagnetic junctions, compared to the TMR$_1$ of the Fe-polariser junctions from Fig.\ref{fig10}(d), and the inset (e) is a zoom around the Fermi level. }{fig12}

The difference with the case of an Fe polariser is mainly in the spin-down transmissions. While in the case of Fe $\Delta_1^\downarrow$ band is absent until above 1.5 eV, this is no longer the case in a fully Mn$_3$Ga FiMTJ and the transmission at the Fermi level for both spin species is dominated by the $\Gamma$ point [see Fig. \ref{fig11}(d)]. It is interesting to examine the energy dependence of the spin-polarised transmission and the TMR effect, which we now define uniquely as the TMR$_1$ from Eq. (\ref{eq:TMR}), because the net and interface spins now are always anti-parallel, so P/AP net moments correspond to P/AP interface spins (Fig. \ref{fig11}b,c), and we only consider Mn-Ga termination for the all Mn$_3$Ga stacks. We find the same dip in the spin-up transmission between 0.15\,eV and 0.45\,eV (see Fig.\ref{fig13}a,b) as in the Fe-based stack, due to the band gap for spin-up in the $\Gamma$-Z direction of Mn$_3$Ga (Fig.\ref{fig06}). This reduction of the spin-up $\Delta_1$ transmission can be seen also in the $\vcr{k}_\perp$-resolved portraits in Fig. \ref{fig13} at 0.2\,eV and it leads to a significant TMR effect in this energy range [Fig. \ref{fig12}(d)]. The TMR effect is, in fact, substantially higher than in the case of the Fe-based MTJ, because now the dominating spin-down transmission in this energy range is less suppressed, compared to the Fe case, where there is no $\Delta_1^\downarrow$ band present at these energies. This is especially valid, and the TMR enhancement is stronger, for the thicker barrier of 5\,MLs. Such structure indeed shows a larger TMR signal in almost all the energy ranges with significant TMR effect compared to the other Mn$_3$Ga-based MTJs studied here. It is likely that further design, using a combination of chemical substitution and strain, could lead to yet higher TMR values.   

\myFig{1}{1}{true}{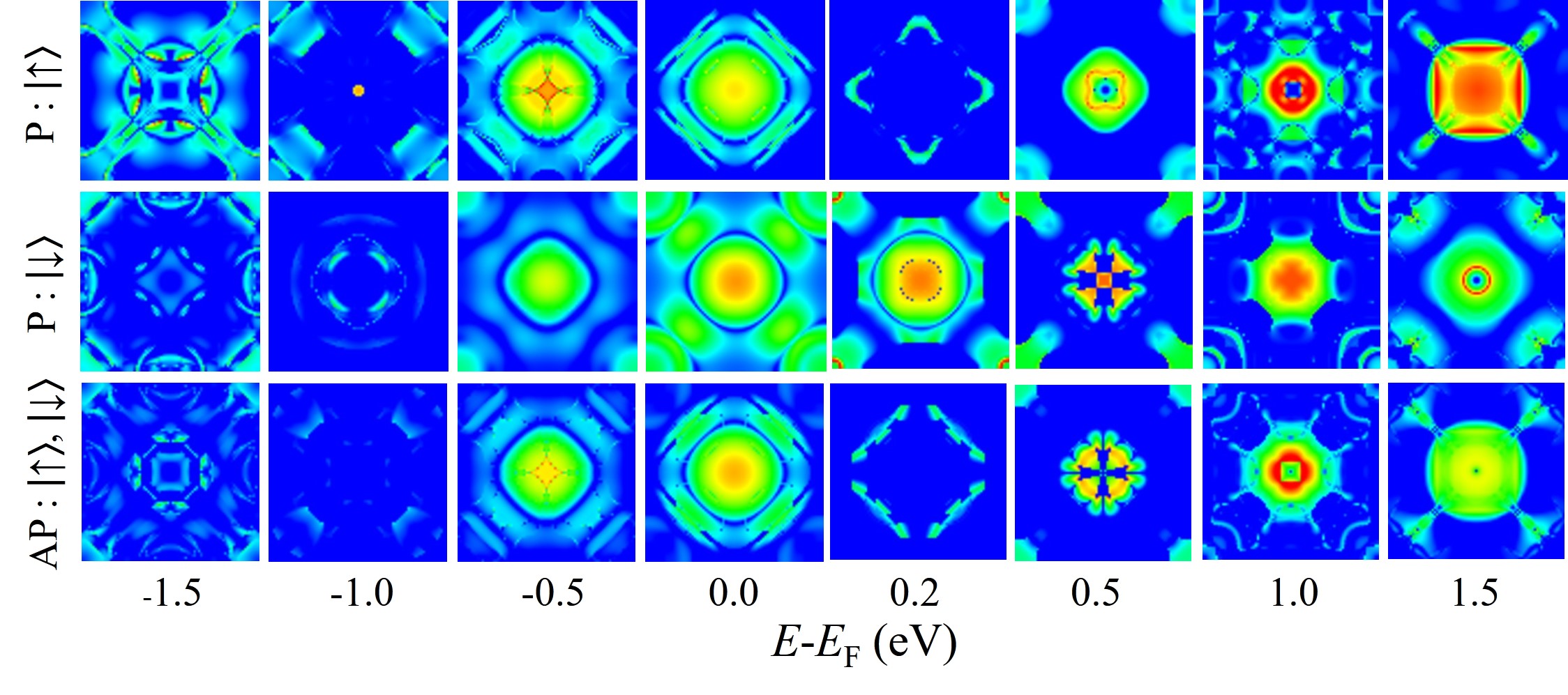}{(Color online) 2D BZ portraits of spin- and energy-resolved transmission. Columns represent the same energy, as indicated at the bottom (not all energy increments between panels are the same). The top two rows are for the two spin-species in the P state, while the bottom row is for both spin-species in the AP state (which are identical for symmetry reasons). The color code for the transmissions is the same as in Fig. \ref{fig11}(d).}{fig13}

In contrast, interestingly, the thinner MgO barrier, i.e. the Mn$_3$Ga/3MgO/Mn$_3$Ga junction, presents an enhancement of the TMR effect at the Fermi level compared to all the other cases [see Fig. \ref{fig12}(e)]. This is due to the enhanced spin-down transmission in the P state, because of symmetry-driven interface resonant states at the Fermi level (similar to the well-known theoretically interface resonances in the minority-spin channel near the Fermi level in Fe/MgO/Fe MTJs\cite{Rungger2009}). Such feature manifests itself in Fig. \ref{fig12}(c) as a peak in the total P-state transmission just above $E_\mathrm{F}$. This leads to a TMR at the Fermi level of about 123 \% for the 3ML-MgO junction, which significantly surpasses the TMR observed in all other Mn$_3$Ga-based junctions we have investigated. In all other energy ranges where we find significant TMR effect, the 5ML-MgO junction shows distinctly higher TMR effect in comparison to the thinner barrier because of the enhanced directional spin filtering. Furthermore, as far as the equilibrium TMR is concerned, there is little difference between the Fe-based and the all-Mn$_3$Ga stack with 5ML MgO [Fig. \ref{fig12}(e)]. This is because the features in the transmission in the $\pm1$\,eV vicinity of $E_\mathrm{F}$ are determined entirely by the electronic structure of the Mn$_3$Ga exhibiting a band-gap between the $\Delta_1^\uparrow$ and $\Delta_5^\uparrow$ bands (which is also rather stable to $c$ constant variations in Mn$_3$Ga, as can be seen from Fig. \ref{fig06}). However, away from equilibrium the anticipated peak in the TMR($V_\mathrm{b}$), produced when this feature enters the bias window, is likely to be higher for the symmetric all-Mn$_3$Ga junction. This is due to the fact that in the case of Fe, the effect of the $\Delta_1^\uparrow$-$\Delta_5^\uparrow$ band gap in Mn$_3$Ga is suppressed in the AP state transmission because of the higher transmission between Fe$^\downarrow$ and Mn$_3$Ga$^\downarrow$ bands [Fig.\ref{fig10}(a)]. This is compared to the lower transmission between Mn$_3$Ga$^\uparrow$ and Mn$_3$Ga$^\downarrow$ [Fig. \ref{fig12}(b) and Fig.\ref{fig13}] in this energy range, where there is a gap in the Mn$_3$Ga$^\uparrow$ band structure.          

\myFig{1}{1}{true}{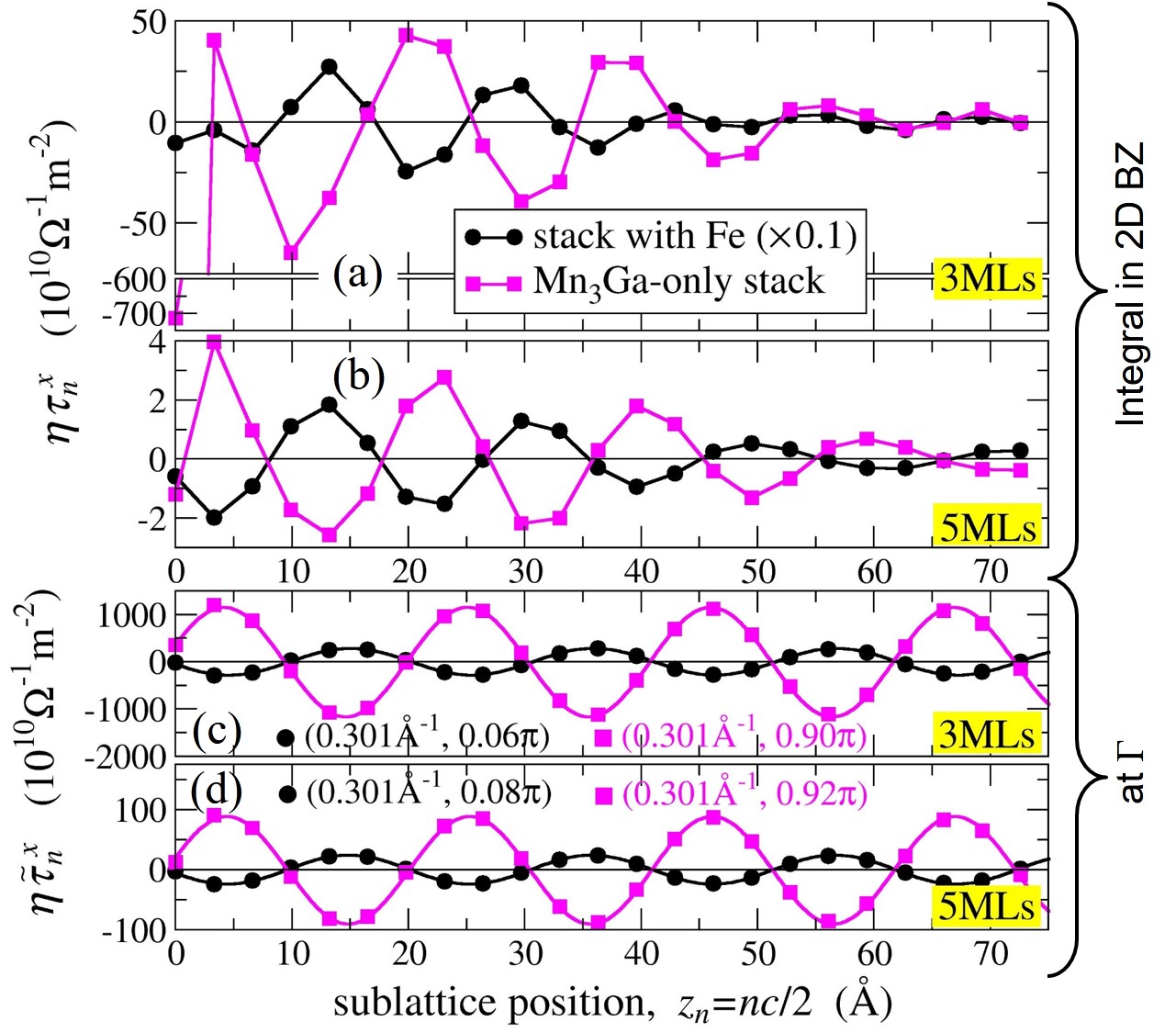}{(Color online) Atomically resolved in-plane STTk for the Mn$_\mathrm{I}$ sublattice in Mn$_3$Ga-only MTJ stacks, compared to previously presented results for the stacks with Fe-polariser for the same thicknesses of 3\,MLs and 5\,MLs of the MgO barriers (as indicated in the panels). (a) and (b) are for the total STTk, while (c) and (d) are at the $\Gamma$ point. Note that there is a broken y-axis in (a) to show the much higher (absolute) value at the interfacial Mn$_\mathrm{I}$ site and there is a scaling factor of $0.1$ for all the data from the Fe-based stacks. }{fig14}

Finally, we look at the STT effect in the all-Mn$_3$Ga based stack and compare that to the Fe-polariser case (see Fig. \ref{fig14}). We find a somewhat suppressed STTk for both barrier thicknesses, which is due to the reduced spin-polarisation of the transmission at the Fermi level, compared to the Fe-based MTJ. The oscillations in the in-plane STTk are still present, but we find that they are further suppressed with respect to the $\Gamma$-point STTk contribution, due to the presence of a larger number of open channels away from $\Gamma$ in the case on an entirely Mn$_3$Ga-based FiMTJ. We also find a significant enhancement of the STTk at the interface of the 3ML-MgO stack, which is due to spin-polarised resonance states at the Mn$_3$Ga-MgO interface of this symmetric-barrier junction. Note that at the first interfacial Mn site we find a nearly seven-fold increase of the in-plane STTk compared to the analogous Fe-polariser junction. In this particular case of the Mn$_3$Ga-only junction we also find a TMR enhancement at the Fermi level, with a theoretical prediction of 123\,\% TMR. Based on the 0\,V transmissions we can anticipate a change of sign in the TMR($V_\mathrm{b}$) below $\pm1$\,V. However, for an accurate analysis of the TMR($V_\mathrm{b}$) self-consistent NEGF+DFT finite bias calculations are required and such will be the subject of another publication. 

\myFig{1.1}{1.1}{true}{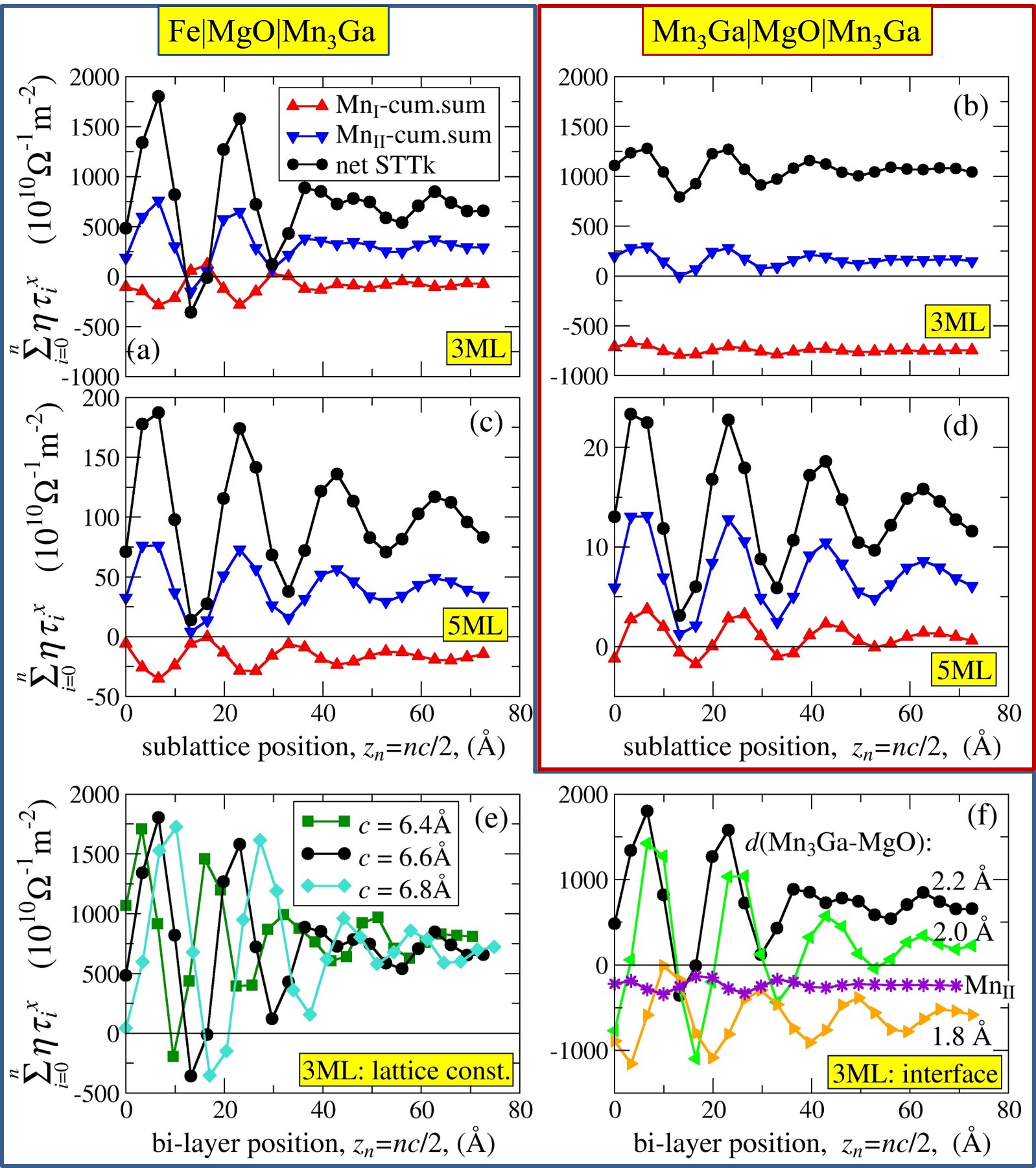}{(Color online) Cumulative sums of atomically-resolved in-plane STTk as a function of the number of bi-layers $n$ (one ML of Mn$_\mathrm{I}$ and one of Mn$_\mathrm{II}$) in Mn$_3$Ga starting from the MgO interface, representing net in-plane STTk on $2n$ MLs of Mn$_3$Ga. In (a-d) show sub-lattice specific quantities, calculated using the results in Fig. \ref{fig14} for the four different junctions (as indicated on the panels), but also not shown there in-plane STTk for the Mn$_\mathrm{II}$ sub-lattice. The net quantity (black circles), representing the torque on the net moment of $2n$ MLs of Mn$_3$Ga, is calculated as $\tau^x_\mathrm{net}(n)=\sum_{i=0}^n \SqB{2\tau^x_i(\mathrm{Mn_{II}})-\tau^x_i(\mathrm{Mn_I})}$ from the two sub-lattice-specific quantities. In bottom panels are $\tau^x_\mathrm{net}(n)$ for other Fe-based junctions, comparing effect of (e) the $c$-constant variation in Mn$_3$Ga (see also Fig. \ref{fig05}) and (f) the variation of the interface spacing $d$ (as in Fig. \ref{fig04}) or Mn$_3$Ga termination on Mn$_\mathrm{II}-$Mn$_\mathrm{II}$ (Fig. \ref{fig03}).  }{fig15}

For these characteristic long-range oscillatory decays of the in-plane STTk in both magnetic sublattices, it is also insightful to look at the net torkance as a function of the number of Mn$_3$Ga layers following the MgO interface. In Fig. \ref{fig15} we compare the Fe-based and the all-Mn$_3$Ga MTJs, for the same interfaces and barriers; we also show results for the different structural parameters in the Fe-based MTJs. We find that in all cases the net in-plane STTk also shows oscillations, but typically there is an offset and a tendency to decay towards a finite (most often positive) value, describing the net STTk in the limit of a thick Mn$_3$Ga layer. For the thinner barrier, the large interface STTk on the first Mn$_\mathrm{I}$-Ga layer, a feature that we attribute to interface resonance states, results in a net STTk, which is even somewhat higher than that in the Fe-based MTJ. Note that the atomically-resolved STTk away from the interface in the all-FiMTJ are almost an order of magnitude smaller than that in the Fe-based MTJs (Fig. \ref{fig14}). This is reflected in the net STTk for the 5 ML junctions [Fig. \ref{fig15}(c,d)], which appear qualitatively identical, albeit scaled by about a factor of 10. It is also interesting to see that despite differences in the phase and frequency of the spatial oscillations as a function of the $c$-constant, the net in-plane STTk for junctions with the same barriers tend to saturate at the same level. Hence, we establish that there is little sensitivity of net in-plane torkance to the lattice parameters of Mn$_3$Ga. What affects the net STTk more significantly is the interlayer distance $d$ and we see that the STTk can change sign with compression of $d$ (although it is likely that such small interface spacings are nonphysical). The interface termination is also important, although our example of Mn$_\mathrm{II}$-Mn$_\mathrm{II}$ termination might not be representative of a real situation, since we have only removed the top Mn$_\mathrm{I}$-Ga layer, but preserved the interface distance and the MgO termination. This case was only included for illustrative purposes to show the effect of changing just one of the interfaces in the junction. Overall, our results suggest that, despite the long-range oscillations, the staggered alignment between the two magnetic sublattices STTk would be giving rise to a net switching torque in the limit of sufficiently thick Mn$_3$Ga slab. Furthermore, in this material, torque modification is possible, using either the thickness of the layer or interface engineering.        

\section{Conclusion}

We report on first principles (SDFT+NEGF) calculations of atomically-resolved spin-transfer torque in a ferrimagnetic tetragonal Mn$_3$Ga in Fe/MgO junction and in an all-ferrimagnetic junction based on Mn$_3$Ga. In a scattering region extending to over 76\,\AA\, of Mn$_3$Ga, we find a long-range oscillatory decay of the STT, both in the linear response (zero-bias) regime and at finite bias. This oscillation is investigated against variations of the material parameters and stack geometry, and found to be persistent. It is quantitatively understood from the bulk electronic structure of Mn$_3$Ga and the spin-filtering properties of the Fe/MgO side of the junction. Spin-transport properties are also compared to the case of an analogous fully Mn$_3$Ga-based FiMTJ stack, which shows similar spatial oscillations and decay rate of the in-plane STT, but usually at a lower by about a factor of 0.1 amplitude. The later junctions have been constructed to be symmetric with odd number of MgO MLs (3 or 5) and are found to be prone to interface states supporting resonant tunneling, especially in the 3 ML case. This leads to enhancements of both the net STT and the TMR effect at the Fermi level, in comparison to the asymmetric Fe-based junctions. We also find a significant enhancement of the out-of-plane (field-like) torque at the interface, which becomes even more pronounced at finite bias. The oscillations in the STTk lead to oscillatory behaviour also of the net in-plane torque as a function of the thickness of Mn$_3$Ga layer, but the net STT stabilises for sufficiently thick layers. The net in-plane torques calculated in Fe/MgO/Mn$_3$Ga junctions saturate at about 750$\times10^{10}\,\Omega^{-1} \mathrm{m}^{-2}$, for 3 MLs of MgO, and to 100$\times10^{10}\,\Omega^{-1} \mathrm{m}^{-2}$, for 5 MLs, which is significantly larger when compared to the net in-plane STT of 2$\times10^{10}\,\Omega^{-1} \mathrm{m}^{-2}$  calculated in analogous Fe/MgO/Fe junctions with 6 MLs MgO.\cite{Rungger2018}     

The bias dependence of the TMR effect in any of the Mn$_3$Ga-based MTJs appears to be largely determined by the band-edges of the spin-up (in our convention) $\Delta_1$-symmetry band in the $z$-direction of tetragonal Mn$_3$Ga. Although, we have not calculated TMR$(V_\mathrm{b})$, we have identified key features in the energy-dependent spin-polarised transmission coefficients, also defining in passing TMR($E$), which would give rise to peaks and sign-changes in TMR$(V_\mathrm{b})$. In the Fe-based structure, we find a modest TMR at equilibrium (up to 50 \%), but a characteristic feature in the transmissions due to a band gap in the $\Gamma$-Z direction for tetragonal Mn$_3$Ga in the energy-range 0.15 - 0.4 eV above the Fermi level, which we expect to result in a peak in the TMR($V_\mathrm{b}$) near 0.5 V. We also anticipate a change in the TMR sign at negative bias (under 1V). Such asymmetric TMR effect, featuring a peak and a sign-change in the range -1 V to 1 V, is in agreement with experimental observations in similar Mn-based FiMTJs\cite{Borisov2016}. 

In the all-Mn$_3$Ga FiMTJs, which we propose, the corresponding island-like feature in the equilibrium TMR($E$) appears enhanced because of the higher spin-down transmission in the AP state, compared to the Fe-based AP case, where the minority spin is missing a $\Delta_1$ band in the Fe electrode, but a lower transmission for both spin species in the AP state of the former junction in this energy range. This enhancement in TMR($E$) is much stronger for the thicker barrier. Interestingly, for the junction with the thinner barrier we find a much more significant with respect to the Fe case TMR effect (some 123 \%) close to equilibrium (0 V) due to surface-state resonant tunneling in this mirror-symmetric junction. Further self-consistent finite bias calculations can elucidate more accurately the anticipated features of the finite bias TMR effect in these systems. However, there are clear indications that, both the all-FiM (still to be demonstrated experimentally) and the one-sided Mn$_3$Ga-based MTJs, which we propose, hold high promise for applications in STT oscillators and memory cells, with their high current-induced torques and encouraging TMR effect.

All authors gratefully acknowledge the joint funding from the Science Foundation Ireland (SFI Grant No. 16/US-C2C/3287) and the National Science Foundation (NSF ERC-TANMS Grant No. 1160504 and NSF-PREM Grant No. DMR-1828019). PS, SS and MS acknowledge funding from the EC H2020 FET-Open project TRANSPIRE (Grant no.  DLV-737038). MS acknowledges a Starting Investigator Research Grant by SFI (Grant No. 18/SIRG/5515). We thank the Irish Centre for High-End Computing (ICHEC) and the Trinity Research IT Centre (TCHPC) for the provision of computational facilities and support.

\bibliography{main}

\begin{thebibliography}{28}%
\makeatletter
\providecommand \@ifxundefined [1]{%
 \@ifx{#1\undefined}
}%
\providecommand \@ifnum [1]{%
 \ifnum #1\expandafter \@firstoftwo
 \else \expandafter \@secondoftwo
 \fi
}%
\providecommand \@ifx [1]{%
 \ifx #1\expandafter \@firstoftwo
 \else \expandafter \@secondoftwo
 \fi
}%
\providecommand \natexlab [1]{#1}%
\providecommand \enquote  [1]{``#1''}%
\providecommand \bibnamefont  [1]{#1}%
\providecommand \bibfnamefont [1]{#1}%
\providecommand \citenamefont [1]{#1}%
\providecommand \href@noop [0]{\@secondoftwo}%
\providecommand \href [0]{\begingroup \@sanitize@url \@href}%
\providecommand \@href[1]{\@@startlink{#1}\@@href}%
\providecommand \@@href[1]{\endgroup#1\@@endlink}%
\providecommand \@sanitize@url [0]{\catcode `\\12\catcode `\$12\catcode
  `\&12\catcode `\#12\catcode `\^12\catcode `\_12\catcode `\%12\relax}%
\providecommand \@@startlink[1]{}%
\providecommand \@@endlink[0]{}%
\providecommand \url  [0]{\begingroup\@sanitize@url \@url }%
\providecommand \@url [1]{\endgroup\@href {#1}{\urlprefix }}%
\providecommand \urlprefix  [0]{URL }%
\providecommand \Eprint [0]{\href }%
\providecommand \doibase [0]{https://doi.org/}%
\providecommand \selectlanguage [0]{\@gobble}%
\providecommand \bibinfo  [0]{\@secondoftwo}%
\providecommand \bibfield  [0]{\@secondoftwo}%
\providecommand \translation [1]{[#1]}%
\providecommand \BibitemOpen [0]{}%
\providecommand \bibitemStop [0]{}%
\providecommand \bibitemNoStop [0]{.\EOS\space}%
\providecommand \EOS [0]{\spacefactor3000\relax}%
\providecommand \BibitemShut  [1]{\csname bibitem#1\endcsname}%
\let\auto@bib@innerbib\@empty
\bibitem [{\citenamefont {Butler}\ \emph {et~al.}(2001)\citenamefont {Butler},
  \citenamefont {Zhang}, \citenamefont {Schulthess},\ and\ \citenamefont
  {MacLaren}}]{Butler2001}%
  \BibitemOpen
  \bibfield  {author} {\bibinfo {author} {\bibfnamefont {W.~H.}\ \bibnamefont
  {Butler}}, \bibinfo {author} {\bibfnamefont {X.-G.}\ \bibnamefont {Zhang}},
  \bibinfo {author} {\bibfnamefont {T.~C.}\ \bibnamefont {Schulthess}},\ and\
  \bibinfo {author} {\bibfnamefont {J.~M.}\ \bibnamefont {MacLaren}},\ }\href
  {https://doi.org/10.1103/PhysRevB.63.054416} {\bibfield  {journal} {\bibinfo
  {journal} {Phys. Rev. B}\ }\textbf {\bibinfo {volume} {63}},\ \bibinfo
  {pages} {054416} (\bibinfo {year} {2001})}\BibitemShut {NoStop}%
\bibitem [{\citenamefont {Parkin}\ \emph {et~al.}(2004)\citenamefont {Parkin},
  \citenamefont {Kaiser}, \citenamefont {Panchula}, \citenamefont {Rice},
  \citenamefont {Hughes}, \citenamefont {Samant},\ and\ \citenamefont
  {Yang}}]{Parkin2004}%
  \BibitemOpen
  \bibfield  {author} {\bibinfo {author} {\bibfnamefont {S.~S.~P.}\
  \bibnamefont {Parkin}}, \bibinfo {author} {\bibfnamefont {C.}~\bibnamefont
  {Kaiser}}, \bibinfo {author} {\bibfnamefont {A.}~\bibnamefont {Panchula}},
  \bibinfo {author} {\bibfnamefont {P.~M.}\ \bibnamefont {Rice}}, \bibinfo
  {author} {\bibfnamefont {B.}~\bibnamefont {Hughes}}, \bibinfo {author}
  {\bibfnamefont {M.}~\bibnamefont {Samant}},\ and\ \bibinfo {author}
  {\bibfnamefont {S.-H.}\ \bibnamefont {Yang}},\ }\href
  {https://doi.org/10.1038/nmat1256} {\bibfield  {journal} {\bibinfo  {journal}
  {Nature Materials}\ }\textbf {\bibinfo {volume} {3}},\ \bibinfo {pages} {862}
  (\bibinfo {year} {2004})}\BibitemShut {NoStop}%
\bibitem [{\citenamefont {Yuasa}\ \emph {et~al.}(2004)\citenamefont {Yuasa},
  \citenamefont {Nagahama}, \citenamefont {Fukushima}, \citenamefont {Suzuki},\
  and\ \citenamefont {Ando}}]{Yuasa2004}%
  \BibitemOpen
  \bibfield  {author} {\bibinfo {author} {\bibfnamefont {S.}~\bibnamefont
  {Yuasa}}, \bibinfo {author} {\bibfnamefont {T.}~\bibnamefont {Nagahama}},
  \bibinfo {author} {\bibfnamefont {A.}~\bibnamefont {Fukushima}}, \bibinfo
  {author} {\bibfnamefont {Y.}~\bibnamefont {Suzuki}},\ and\ \bibinfo {author}
  {\bibfnamefont {K.}~\bibnamefont {Ando}},\ }\href
  {https://doi.org/10.1038/nmat1257} {\bibfield  {journal} {\bibinfo  {journal}
  {Nature Materials}\ }\textbf {\bibinfo {volume} {3}},\ \bibinfo {pages} {868}
  (\bibinfo {year} {2004})}\BibitemShut {NoStop}%
\bibitem [{\citenamefont {Butler}(2008)}]{Butler2008}%
  \BibitemOpen
  \bibfield  {author} {\bibinfo {author} {\bibfnamefont {W.~H.}\ \bibnamefont
  {Butler}},\ }\href {https://doi.org/10.1088/1468-6996/9/1/014106} {\bibfield
  {journal} {\bibinfo  {journal} {Sci. Tech. Adv. Mater.}\ }\textbf {\bibinfo
  {volume} {9}},\ \bibinfo {pages} {014106} (\bibinfo {year}
  {2008})}\BibitemShut {NoStop}%
\bibitem [{\citenamefont {Rungger}\ \emph {et~al.}(2009)\citenamefont
  {Rungger}, \citenamefont {Mryasov},\ and\ \citenamefont
  {Sanvito}}]{Rungger2009}%
  \BibitemOpen
  \bibfield  {author} {\bibinfo {author} {\bibfnamefont {I.}~\bibnamefont
  {Rungger}}, \bibinfo {author} {\bibfnamefont {O.}~\bibnamefont {Mryasov}},\
  and\ \bibinfo {author} {\bibfnamefont {S.}~\bibnamefont {Sanvito}},\ }\href
  {https://doi.org/10.1103/PhysRevB.79.094414} {\bibfield  {journal} {\bibinfo
  {journal} {Phys. Rev. B}\ }\textbf {\bibinfo {volume} {79}},\ \bibinfo
  {pages} {2} (\bibinfo {year} {2009})}\BibitemShut {NoStop}%
\bibitem [{\citenamefont {Rungger}\ \emph {et~al.}(2018)\citenamefont
  {Rungger}, \citenamefont {Droghetti},\ and\ \citenamefont
  {Stamenova}}]{Rungger2018}%
  \BibitemOpen
  \bibfield  {author} {\bibinfo {author} {\bibfnamefont {I.}~\bibnamefont
  {Rungger}}, \bibinfo {author} {\bibfnamefont {A.}~\bibnamefont {Droghetti}},\
  and\ \bibinfo {author} {\bibfnamefont {M.}~\bibnamefont {Stamenova}},\
  }\bibinfo {title} {Non-equilibrium green's function methods for spin
  transport and dynamics},\ in\ \href
  {https://doi.org/10.1007/978-3-319-42913-7_75-1} {\emph {\bibinfo {booktitle}
  {Handbook of Materials Modeling : Methods: Theory and Modeling}}},\ \bibinfo
  {editor} {edited by\ \bibinfo {editor} {\bibfnamefont {W.}~\bibnamefont
  {Andreoni}}\ and\ \bibinfo {editor} {\bibfnamefont {S.}~\bibnamefont {Yip}}}\
  (\bibinfo  {publisher} {Springer International Publishing},\ \bibinfo
  {address} {Cham},\ \bibinfo {year} {2018})\ pp.\ \bibinfo {pages}
  {957--983}\BibitemShut {NoStop}%
\bibitem [{\citenamefont {Houssameddine}\ \emph {et~al.}(2007)\citenamefont
  {Houssameddine}, \citenamefont {Ebels}, \citenamefont {Dela{\"{e}}t},
  \citenamefont {Rodmacq}, \citenamefont {Firastrau}, \citenamefont
  {Ponthenier}, \citenamefont {Brunet}, \citenamefont {Thirion}, \citenamefont
  {Michel}, \citenamefont {Prejbeanu-Buda}, \citenamefont {Cyrille},
  \citenamefont {Redon},\ and\ \citenamefont {Dieny}}]{Houssameddine2007}%
  \BibitemOpen
  \bibfield  {author} {\bibinfo {author} {\bibfnamefont {D.}~\bibnamefont
  {Houssameddine}}, \bibinfo {author} {\bibfnamefont {U.}~\bibnamefont
  {Ebels}}, \bibinfo {author} {\bibfnamefont {B.}~\bibnamefont {Dela{\"{e}}t}},
  \bibinfo {author} {\bibfnamefont {B.}~\bibnamefont {Rodmacq}}, \bibinfo
  {author} {\bibfnamefont {I.}~\bibnamefont {Firastrau}}, \bibinfo {author}
  {\bibfnamefont {F.}~\bibnamefont {Ponthenier}}, \bibinfo {author}
  {\bibfnamefont {M.}~\bibnamefont {Brunet}}, \bibinfo {author} {\bibfnamefont
  {C.}~\bibnamefont {Thirion}}, \bibinfo {author} {\bibfnamefont {J.~P.}\
  \bibnamefont {Michel}}, \bibinfo {author} {\bibfnamefont {L.}~\bibnamefont
  {Prejbeanu-Buda}}, \bibinfo {author} {\bibfnamefont {M.~C.}\ \bibnamefont
  {Cyrille}}, \bibinfo {author} {\bibfnamefont {O.}~\bibnamefont {Redon}},\
  and\ \bibinfo {author} {\bibfnamefont {B.}~\bibnamefont {Dieny}},\ }\href
  {https://doi.org/10.1038/nmat1905} {\bibfield  {journal} {\bibinfo  {journal}
  {Nature Materials}\ }\textbf {\bibinfo {volume} {6}},\ \bibinfo {pages} {447}
  (\bibinfo {year} {2007})}\BibitemShut {NoStop}%
\bibitem [{\citenamefont {Heindl}\ and\ \citenamefont
  {Rippard}(2019)}]{Heindl2019}%
  \BibitemOpen
  \bibfield  {author} {\bibinfo {author} {\bibfnamefont {R.}~\bibnamefont
  {Heindl}}\ and\ \bibinfo {author} {\bibfnamefont {W.~H.}\ \bibnamefont
  {Rippard}},\ }\href {https://doi.org/10.1149/1.3119524} {\bibfield  {journal}
  {\bibinfo  {journal} {ECS Transactions}\ }\textbf {\bibinfo {volume} {19}},\
  \bibinfo {pages} {21} (\bibinfo {year} {2019})}\BibitemShut {NoStop}%
\bibitem [{\citenamefont {Hirohata}\ \emph {et~al.}(2020)\citenamefont
  {Hirohata}, \citenamefont {Yamada}, \citenamefont {Nakatani}, \citenamefont
  {Prejbeanu}, \citenamefont {Di{\'{e}}ny}, \citenamefont {Pirro},\ and\
  \citenamefont {Hillebrands}}]{Hirohata2020}%
  \BibitemOpen
  \bibfield  {author} {\bibinfo {author} {\bibfnamefont {A.}~\bibnamefont
  {Hirohata}}, \bibinfo {author} {\bibfnamefont {K.}~\bibnamefont {Yamada}},
  \bibinfo {author} {\bibfnamefont {Y.}~\bibnamefont {Nakatani}}, \bibinfo
  {author} {\bibfnamefont {L.}~\bibnamefont {Prejbeanu}}, \bibinfo {author}
  {\bibfnamefont {B.}~\bibnamefont {Di{\'{e}}ny}}, \bibinfo {author}
  {\bibfnamefont {P.}~\bibnamefont {Pirro}},\ and\ \bibinfo {author}
  {\bibfnamefont {B.}~\bibnamefont {Hillebrands}},\ }\href
  {http://doi.org/10.1016/j.jmmm.2020.166711} {\bibfield  {journal} {\bibinfo
  {journal} {J. Magn. Magn. Materials}\ }\textbf {\bibinfo {volume} {509}}
  (\bibinfo {year} {2020})}\BibitemShut {NoStop}%
\bibitem [{\citenamefont {Mizukami}\ \emph {et~al.}(2011)\citenamefont
  {Mizukami}, \citenamefont {Wu}, \citenamefont {Sakuma}, \citenamefont
  {Walowski}, \citenamefont {Watanabe}, \citenamefont {Kubota}, \citenamefont
  {Zhang}, \citenamefont {Naganuma}, \citenamefont {Oogane}, \citenamefont
  {Ando},\ and\ \citenamefont {Miyazaki}}]{Mizukami2011}%
  \BibitemOpen
  \bibfield  {author} {\bibinfo {author} {\bibfnamefont {S.}~\bibnamefont
  {Mizukami}}, \bibinfo {author} {\bibfnamefont {F.}~\bibnamefont {Wu}},
  \bibinfo {author} {\bibfnamefont {A.}~\bibnamefont {Sakuma}}, \bibinfo
  {author} {\bibfnamefont {J.}~\bibnamefont {Walowski}}, \bibinfo {author}
  {\bibfnamefont {D.}~\bibnamefont {Watanabe}}, \bibinfo {author}
  {\bibfnamefont {T.}~\bibnamefont {Kubota}}, \bibinfo {author} {\bibfnamefont
  {X.}~\bibnamefont {Zhang}}, \bibinfo {author} {\bibfnamefont
  {H.}~\bibnamefont {Naganuma}}, \bibinfo {author} {\bibfnamefont
  {M.}~\bibnamefont {Oogane}}, \bibinfo {author} {\bibfnamefont
  {Y.}~\bibnamefont {Ando}},\ and\ \bibinfo {author} {\bibfnamefont
  {T.}~\bibnamefont {Miyazaki}},\ }\href
  {https://link.aps.org/doi/10.1103/PhysRevLett.106.117201} {\bibfield
  {journal} {\bibinfo  {journal} {Phys. Rev. Lett.}\ }\textbf {\bibinfo
  {volume} {106}},\ \bibinfo {pages} {117201} (\bibinfo {year}
  {2011})}\BibitemShut {NoStop}%
\bibitem [{\citenamefont {Rode}\ \emph {et~al.}(2013)\citenamefont {Rode},
  \citenamefont {Baadji}, \citenamefont {Betto}, \citenamefont {Lau},
  \citenamefont {Kurt}, \citenamefont {Venkatesan}, \citenamefont {Stamenov},
  \citenamefont {Sanvito}, \citenamefont {Coey}, \citenamefont {Fonda},
  \citenamefont {Otero}, \citenamefont {Choueikani}, \citenamefont {Ohresser},
  \citenamefont {Porcher},\ and\ \citenamefont {Andr\'e}}]{Rode2013}%
  \BibitemOpen
  \bibfield  {author} {\bibinfo {author} {\bibfnamefont {K.}~\bibnamefont
  {Rode}}, \bibinfo {author} {\bibfnamefont {N.}~\bibnamefont {Baadji}},
  \bibinfo {author} {\bibfnamefont {D.}~\bibnamefont {Betto}}, \bibinfo
  {author} {\bibfnamefont {Y.~C.}\ \bibnamefont {Lau}}, \bibinfo {author}
  {\bibfnamefont {H.}~\bibnamefont {Kurt}}, \bibinfo {author} {\bibfnamefont
  {M.}~\bibnamefont {Venkatesan}}, \bibinfo {author} {\bibfnamefont
  {P.}~\bibnamefont {Stamenov}}, \bibinfo {author} {\bibfnamefont
  {S.}~\bibnamefont {Sanvito}}, \bibinfo {author} {\bibfnamefont {J.~M.~D.}\
  \bibnamefont {Coey}}, \bibinfo {author} {\bibfnamefont {E.}~\bibnamefont
  {Fonda}}, \bibinfo {author} {\bibfnamefont {E.}~\bibnamefont {Otero}},
  \bibinfo {author} {\bibfnamefont {F.}~\bibnamefont {Choueikani}}, \bibinfo
  {author} {\bibfnamefont {P.}~\bibnamefont {Ohresser}}, \bibinfo {author}
  {\bibfnamefont {F.}~\bibnamefont {Porcher}},\ and\ \bibinfo {author}
  {\bibfnamefont {G.}~\bibnamefont {Andr\'e}},\ }\href
  {https://doi.org/10.1103/PhysRevB.87.184429} {\bibfield  {journal} {\bibinfo
  {journal} {Phys. Rev. B}\ }\textbf {\bibinfo {volume} {87}},\ \bibinfo
  {pages} {184429} (\bibinfo {year} {2013})}\BibitemShut {NoStop}%
\bibitem [{\citenamefont {Finley}\ \emph {et~al.}(2019)\citenamefont {Finley},
  \citenamefont {Lee}, \citenamefont {Huang},\ and\ \citenamefont
  {Liu}}]{Finley2018}%
  \BibitemOpen
  \bibfield  {author} {\bibinfo {author} {\bibfnamefont {J.}~\bibnamefont
  {Finley}}, \bibinfo {author} {\bibfnamefont {C.~H.}\ \bibnamefont {Lee}},
  \bibinfo {author} {\bibfnamefont {P.~Y.}\ \bibnamefont {Huang}},\ and\
  \bibinfo {author} {\bibfnamefont {L.}~\bibnamefont {Liu}},\ }\href
  {https://doi.org/10.1002/adma.201805361} {\bibfield  {journal} {\bibinfo
  {journal} {Adv. Mater.}\ }\textbf {\bibinfo {volume} {31}},\ \bibinfo {pages}
  {1805361} (\bibinfo {year} {2019})}\BibitemShut {NoStop}%
\bibitem [{\citenamefont {Ilyakov}\ \emph {et~al.}(2019)\citenamefont
  {Ilyakov}, \citenamefont {Awari}, \citenamefont {Kovalev}, \citenamefont
  {Fowley}, \citenamefont {Rode}, \citenamefont {Stamenov}, \citenamefont
  {Lau}, \citenamefont {Betto}, \citenamefont {Thiyagarajah}, \citenamefont
  {Green}, \citenamefont {Yildirim}, \citenamefont {Lindner}, \citenamefont
  {Fassbender}, \citenamefont {Coey}, \citenamefont {Deac},\ and\ \citenamefont
  {Gensch}}]{Ilyakov2019}%
  \BibitemOpen
  \bibfield  {author} {\bibinfo {author} {\bibfnamefont {I.}~\bibnamefont
  {Ilyakov}}, \bibinfo {author} {\bibfnamefont {N.}~\bibnamefont {Awari}},
  \bibinfo {author} {\bibfnamefont {S.}~\bibnamefont {Kovalev}}, \bibinfo
  {author} {\bibfnamefont {C.}~\bibnamefont {Fowley}}, \bibinfo {author}
  {\bibfnamefont {K.}~\bibnamefont {Rode}}, \bibinfo {author} {\bibfnamefont
  {P.}~\bibnamefont {Stamenov}}, \bibinfo {author} {\bibfnamefont {Y.-C.}\
  \bibnamefont {Lau}}, \bibinfo {author} {\bibfnamefont {D.}~\bibnamefont
  {Betto}}, \bibinfo {author} {\bibfnamefont {N.}~\bibnamefont {Thiyagarajah}},
  \bibinfo {author} {\bibfnamefont {B.}~\bibnamefont {Green}}, \bibinfo
  {author} {\bibfnamefont {O.}~\bibnamefont {Yildirim}}, \bibinfo {author}
  {\bibfnamefont {J.}~\bibnamefont {Lindner}}, \bibinfo {author} {\bibfnamefont
  {J.}~\bibnamefont {Fassbender}}, \bibinfo {author} {\bibfnamefont
  {M.}~\bibnamefont {Coey}}, \bibinfo {author} {\bibfnamefont {A.}~\bibnamefont
  {Deac}},\ and\ \bibinfo {author} {\bibfnamefont {M.}~\bibnamefont {Gensch}},\
  }\href {https://doi.org/10.1364/CLEO_SI.2019.STu4F.6} {\bibfield  {journal}
  {\bibinfo  {journal} {Conference on Lasers and Electro-Optics}\ ,\ \bibinfo
  {pages} {STu4F.6}} (\bibinfo {year} {2019})}\BibitemShut {NoStop}%
\bibitem [{\citenamefont {Awari}\ \emph {et~al.}(2016)\citenamefont {Awari},
  \citenamefont {Kovalev}, \citenamefont {Fowley}, \citenamefont {Rode},
  \citenamefont {Gallardo}, \citenamefont {Lau}, \citenamefont {Betto},
  \citenamefont {Thiyagarajah}, \citenamefont {Green}, \citenamefont
  {Yildirim}, \citenamefont {Lindner}, \citenamefont {Fassbender},
  \citenamefont {Coey}, \citenamefont {Deac},\ and\ \citenamefont
  {Gensch}}]{Awari2016}%
  \BibitemOpen
  \bibfield  {author} {\bibinfo {author} {\bibfnamefont {N.}~\bibnamefont
  {Awari}}, \bibinfo {author} {\bibfnamefont {S.}~\bibnamefont {Kovalev}},
  \bibinfo {author} {\bibfnamefont {C.}~\bibnamefont {Fowley}}, \bibinfo
  {author} {\bibfnamefont {K.}~\bibnamefont {Rode}}, \bibinfo {author}
  {\bibfnamefont {R.~A.}\ \bibnamefont {Gallardo}}, \bibinfo {author}
  {\bibfnamefont {Y.~C.}\ \bibnamefont {Lau}}, \bibinfo {author} {\bibfnamefont
  {D.}~\bibnamefont {Betto}}, \bibinfo {author} {\bibfnamefont
  {N.}~\bibnamefont {Thiyagarajah}}, \bibinfo {author} {\bibfnamefont
  {B.}~\bibnamefont {Green}}, \bibinfo {author} {\bibfnamefont
  {O.}~\bibnamefont {Yildirim}}, \bibinfo {author} {\bibfnamefont
  {J.}~\bibnamefont {Lindner}}, \bibinfo {author} {\bibfnamefont
  {J.}~\bibnamefont {Fassbender}}, \bibinfo {author} {\bibfnamefont {J.~M.}\
  \bibnamefont {Coey}}, \bibinfo {author} {\bibfnamefont {A.~M.}\ \bibnamefont
  {Deac}},\ and\ \bibinfo {author} {\bibfnamefont {M.}~\bibnamefont {Gensch}},\
  }\href {http://doi.org/10.1063/1.4958855} {\bibfield  {journal} {\bibinfo
  {journal} {Appl. Phys. Lett.}\ }\textbf {\bibinfo {volume} {109}} (\bibinfo
  {year} {2016})}\BibitemShut {NoStop}%
\bibitem [{\citenamefont {Rocha}\ and\ \citenamefont {Sanvito}(2004)}]{Alex04}%
  \BibitemOpen
  \bibfield  {author} {\bibinfo {author} {\bibfnamefont {A.~R.}\ \bibnamefont
  {Rocha}}\ and\ \bibinfo {author} {\bibfnamefont {S.}~\bibnamefont
  {Sanvito}},\ }\href {http://link.aps.org/doi/10.1103/PhysRevB.70.094406}
  {\bibfield  {journal} {\bibinfo  {journal} {Phys. Rev. B}\ }\textbf {\bibinfo
  {volume} {70}},\ \bibinfo {pages} {094406} (\bibinfo {year}
  {2004})}\BibitemShut {NoStop}%
\bibitem [{\citenamefont {Perdew}\ and\ \citenamefont
  {Zunger}(1981)}]{Perdew1981}%
  \BibitemOpen
  \bibfield  {author} {\bibinfo {author} {\bibfnamefont {J.~P.}\ \bibnamefont
  {Perdew}}\ and\ \bibinfo {author} {\bibfnamefont {A.}~\bibnamefont
  {Zunger}},\ }\href
  {https://journals.aps.org/prb/abstract/10.1103/PhysRevB.23.5048} {\bibfield
  {journal} {\bibinfo  {journal} {Phys. Rev. B}\ }\textbf {\bibinfo {volume}
  {23}},\ \bibinfo {pages} {5075} (\bibinfo {year} {1981})}\BibitemShut
  {NoStop}%
\bibitem [{\citenamefont {{\v{Z}}ic}\ \emph {et~al.}(2016)\citenamefont
  {{\v{Z}}ic}, \citenamefont {Rode}, \citenamefont {Thiyagarajah},
  \citenamefont {Lau}, \citenamefont {Betto}, \citenamefont {Coey},
  \citenamefont {Sanvito}, \citenamefont {O'Shea}, \citenamefont {Ferguson},
  \citenamefont {MacLaren},\ and\ \citenamefont {Archer}}]{Zic2016}%
  \BibitemOpen
  \bibfield  {author} {\bibinfo {author} {\bibfnamefont {M.}~\bibnamefont
  {{\v{Z}}ic}}, \bibinfo {author} {\bibfnamefont {K.}~\bibnamefont {Rode}},
  \bibinfo {author} {\bibfnamefont {N.}~\bibnamefont {Thiyagarajah}}, \bibinfo
  {author} {\bibfnamefont {Y.-C.}\ \bibnamefont {Lau}}, \bibinfo {author}
  {\bibfnamefont {D.}~\bibnamefont {Betto}}, \bibinfo {author} {\bibfnamefont
  {J.~M.~D.}\ \bibnamefont {Coey}}, \bibinfo {author} {\bibfnamefont
  {S.}~\bibnamefont {Sanvito}}, \bibinfo {author} {\bibfnamefont {K.~J.}\
  \bibnamefont {O'Shea}}, \bibinfo {author} {\bibfnamefont {C.~A.}\
  \bibnamefont {Ferguson}}, \bibinfo {author} {\bibfnamefont {D.~A.}\
  \bibnamefont {MacLaren}},\ and\ \bibinfo {author} {\bibfnamefont
  {T.}~\bibnamefont {Archer}},\ }\href
  {https://doi.org/10.1103/PhysRevB.93.140202} {\bibfield  {journal} {\bibinfo
  {journal} {Phys. Rev. B}\ }\textbf {\bibinfo {volume} {93}},\ \bibinfo
  {pages} {140202} (\bibinfo {year} {2016})}\BibitemShut {NoStop}%
\bibitem [{Note1()}]{Note1}%
  \BibitemOpen
  \bibinfo {note} {From the International Centre for Diffraction Data, \protect
  \url {https://www.icdd.com/}.}\BibitemShut {Stop}%
\bibitem [{\citenamefont {Sun}\ \emph {et~al.}(2020)\citenamefont {Sun},
  \citenamefont {Kwon}, \citenamefont {Stamenova}, \citenamefont {Sanvito},\
  and\ \citenamefont {Kioussis}}]{Sun2020}%
  \BibitemOpen
  \bibfield  {author} {\bibinfo {author} {\bibfnamefont {Q.}~\bibnamefont
  {Sun}}, \bibinfo {author} {\bibfnamefont {S.}~\bibnamefont {Kwon}}, \bibinfo
  {author} {\bibfnamefont {M.}~\bibnamefont {Stamenova}}, \bibinfo {author}
  {\bibfnamefont {S.}~\bibnamefont {Sanvito}},\ and\ \bibinfo {author}
  {\bibfnamefont {N.}~\bibnamefont {Kioussis}},\ }\href
  {https://doi.org/10.1103/PhysRevB.101.134419} {\bibfield  {journal} {\bibinfo
   {journal} {Phys. Rev. B}\ }\textbf {\bibinfo {volume} {101}},\ \bibinfo
  {pages} {134419} (\bibinfo {year} {2020})}\BibitemShut {NoStop}%
\bibitem [{Note2()}]{Note2}%
  \BibitemOpen
  \bibinfo {note} {The junctions targeted for experimental comparisons are
  primarily based on ultra-thin CoFeB polarizers, which indeed have a
  well-developed perpendicular anisotropy, provided by the spin-orbit coupling
  at the interface of Fe and MgO. Here we neglect the small differences in
  polarisation and anisotropy between Fe and CoFe.}\BibitemShut {Stop}%
\bibitem [{\citenamefont {Borisov}\ \emph {et~al.}(2016)\citenamefont
  {Borisov}, \citenamefont {Betto}, \citenamefont {Lau}, \citenamefont
  {Fowley}, \citenamefont {Titova}, \citenamefont {Thiyagarajah}, \citenamefont
  {Atcheson}, \citenamefont {Lindner}, \citenamefont {Deac}, \citenamefont
  {Coey}, \citenamefont {Stamenov},\ and\ \citenamefont {Rode}}]{Borisov2016}%
  \BibitemOpen
  \bibfield  {author} {\bibinfo {author} {\bibfnamefont {K.}~\bibnamefont
  {Borisov}}, \bibinfo {author} {\bibfnamefont {D.}~\bibnamefont {Betto}},
  \bibinfo {author} {\bibfnamefont {Y.~C.}\ \bibnamefont {Lau}}, \bibinfo
  {author} {\bibfnamefont {C.}~\bibnamefont {Fowley}}, \bibinfo {author}
  {\bibfnamefont {A.}~\bibnamefont {Titova}}, \bibinfo {author} {\bibfnamefont
  {N.}~\bibnamefont {Thiyagarajah}}, \bibinfo {author} {\bibfnamefont
  {G.}~\bibnamefont {Atcheson}}, \bibinfo {author} {\bibfnamefont
  {J.}~\bibnamefont {Lindner}}, \bibinfo {author} {\bibfnamefont {A.~M.}\
  \bibnamefont {Deac}}, \bibinfo {author} {\bibfnamefont {J.~M.}\ \bibnamefont
  {Coey}}, \bibinfo {author} {\bibfnamefont {P.}~\bibnamefont {Stamenov}},\
  and\ \bibinfo {author} {\bibfnamefont {K.}~\bibnamefont {Rode}},\ }\href
  {https://aip.scitation.org/doi/10.1063/1.4948934} {\bibfield  {journal}
  {\bibinfo  {journal} {Appl. Phys. Lett.}\ }\textbf {\bibinfo {volume} {108}}
  (\bibinfo {year} {2016})}\BibitemShut {NoStop}%
\bibitem [{\citenamefont {Stamenova}\ \emph {et~al.}(2017)\citenamefont
  {Stamenova}, \citenamefont {Mohebbi}, \citenamefont {Seyed-Yazdi},
  \citenamefont {Rungger},\ and\ \citenamefont {Sanvito}}]{Stamenova2016}%
  \BibitemOpen
  \bibfield  {author} {\bibinfo {author} {\bibfnamefont {M.}~\bibnamefont
  {Stamenova}}, \bibinfo {author} {\bibfnamefont {R.}~\bibnamefont {Mohebbi}},
  \bibinfo {author} {\bibfnamefont {J.}~\bibnamefont {Seyed-Yazdi}}, \bibinfo
  {author} {\bibfnamefont {I.}~\bibnamefont {Rungger}},\ and\ \bibinfo {author}
  {\bibfnamefont {S.}~\bibnamefont {Sanvito}},\ }\href
  {http://link.aps.org/doi/10.1103/PhysRevB.95.060403} {\bibfield  {journal}
  {\bibinfo  {journal} {Phys. Rev. B}\ }\textbf {\bibinfo {volume} {95}},\
  \bibinfo {pages} {060403} (\bibinfo {year} {2017})}\BibitemShut {NoStop}%
\bibitem [{\citenamefont {Soler}\ \emph {et~al.}(2002)\citenamefont {Soler},
  \citenamefont {Artacho}, \citenamefont {Gale}, \citenamefont {Garc{\'{i}}a},
  \citenamefont {Junquera}, \citenamefont {Ordej{\'{o}}n},\ and\ \citenamefont
  {S{\'{a}}nchez-Portal}}]{Soler2002}%
  \BibitemOpen
  \bibfield  {author} {\bibinfo {author} {\bibfnamefont {J.~M.}\ \bibnamefont
  {Soler}}, \bibinfo {author} {\bibfnamefont {E.}~\bibnamefont {Artacho}},
  \bibinfo {author} {\bibfnamefont {J.~D.}\ \bibnamefont {Gale}}, \bibinfo
  {author} {\bibfnamefont {A.}~\bibnamefont {Garc{\'{i}}a}}, \bibinfo {author}
  {\bibfnamefont {J.}~\bibnamefont {Junquera}}, \bibinfo {author}
  {\bibfnamefont {P.}~\bibnamefont {Ordej{\'{o}}n}},\ and\ \bibinfo {author}
  {\bibfnamefont {D.}~\bibnamefont {S{\'{a}}nchez-Portal}},\ }\href
  {https://doi.org/10.1088/0953-8984/14/11/302} {\bibfield  {journal} {\bibinfo
   {journal} {J. Phys.: Condens. Matt.}\ }\textbf {\bibinfo {volume} {14}},\
  \bibinfo {pages} {2745} (\bibinfo {year} {2002})}\BibitemShut {NoStop}%
\bibitem [{Note3()}]{Note3}%
  \BibitemOpen
  \bibinfo {note} {Note that we use 'spins' and 'magnetic moments'
  interchangeably throughout the paper.}\BibitemShut {Stop}%
\bibitem [{\citenamefont {Stiles}\ and\ \citenamefont
  {Zangwill}(2002)}]{Stiles2002}%
  \BibitemOpen
  \bibfield  {author} {\bibinfo {author} {\bibfnamefont {M.~D.}\ \bibnamefont
  {Stiles}}\ and\ \bibinfo {author} {\bibfnamefont {A.}~\bibnamefont
  {Zangwill}},\ }\href {https://doi.org/10.1103/PhysRevB.66.014407} {\bibfield
  {journal} {\bibinfo  {journal} {Phys. Rev. B}\ }\textbf {\bibinfo {volume}
  {66}},\ \bibinfo {pages} {144071} (\bibinfo {year} {2002})}\BibitemShut
  {NoStop}%
\bibitem [{\citenamefont {Galante}\ \emph {et~al.}(2019)\citenamefont
  {Galante}, \citenamefont {Ellis},\ and\ \citenamefont
  {Sanvito}}]{Galante2019}%
  \BibitemOpen
  \bibfield  {author} {\bibinfo {author} {\bibfnamefont {M.}~\bibnamefont
  {Galante}}, \bibinfo {author} {\bibfnamefont {M.~O.}\ \bibnamefont {Ellis}},\
  and\ \bibinfo {author} {\bibfnamefont {S.}~\bibnamefont {Sanvito}},\ }\href
  {https://doi.org/10.1103/PhysRevB.99.014401} {\bibfield  {journal} {\bibinfo
  {journal} {Phys. Rev. B}\ }\textbf {\bibinfo {volume} {99}},\ \bibinfo
  {pages} {1} (\bibinfo {year} {2019})}\BibitemShut {NoStop}%
\bibitem [{\citenamefont {Betto}\ \emph {et~al.}(2015)\citenamefont {Betto},
  \citenamefont {Thiyagarajah}, \citenamefont {Lau}, \citenamefont
  {Piamonteze}, \citenamefont {Arrio}, \citenamefont {Stamenov}, \citenamefont
  {Coey},\ and\ \citenamefont {Rode}}]{Betto2015}%
  \BibitemOpen
  \bibfield  {author} {\bibinfo {author} {\bibfnamefont {D.}~\bibnamefont
  {Betto}}, \bibinfo {author} {\bibfnamefont {N.}~\bibnamefont {Thiyagarajah}},
  \bibinfo {author} {\bibfnamefont {Y.~C.}\ \bibnamefont {Lau}}, \bibinfo
  {author} {\bibfnamefont {C.}~\bibnamefont {Piamonteze}}, \bibinfo {author}
  {\bibfnamefont {M.~A.}\ \bibnamefont {Arrio}}, \bibinfo {author}
  {\bibfnamefont {P.}~\bibnamefont {Stamenov}}, \bibinfo {author}
  {\bibfnamefont {J.~M.}\ \bibnamefont {Coey}},\ and\ \bibinfo {author}
  {\bibfnamefont {K.}~\bibnamefont {Rode}},\ }\href
  {http://doi.org/10.1103/PhysRevB.91.094410} {\bibfield  {journal} {\bibinfo
  {journal} {Phys. Rev. B}\ }\textbf {\bibinfo {volume} {91}},\ \bibinfo
  {pages} {1} (\bibinfo {year} {2015})}\BibitemShut {NoStop}%
\bibitem [{Note4()}]{Note4}%
  \BibitemOpen
  \bibinfo {note} {It is difficult to quantify the decay rate for this system
  size and the 'rectangular wave packet' fit from Fig. \ref {fig01} does not
  show a difference in the dispersions on the wave number between the two
  barrier thicknesses.}\BibitemShut {Stop}%
\end{thebibliography}%

\end{document}